\documentclass[
	fontsize=11pt,
	paper=a4,
    abstract=false,
    fleqn,
]{scrartcl}

\usepackage[T1]{fontenc}
\usepackage{lmodern}
\usepackage[utf8]{inputenc}
\usepackage[main=english,ngerman]{babel}
\usepackage[fleqn]{amsmath}
\usepackage{amsfonts}
\usepackage{amssymb}
\usepackage{amsthm}
\usepackage{bm}
\usepackage{siunitx}
\usepackage{scrlayer-scrpage}
\usepackage{geometry}
\usepackage{scrlayer-notecolumn}
\usepackage{graphicx}
\usepackage{xcolor}
\usepackage{csquotes}
\usepackage[shortlabels]{enumitem}
\usepackage{booktabs}
\usepackage{longtable}
\usepackage[section]{placeins}
\usepackage[ruled,vlined,linesnumbered]{algorithm2e}
\usepackage{lipsum}
\usepackage{lastpage}
\usepackage[style=numeric,giveninits=true,isbn=false,eprint=false,url=false]{biblatex}
\usepackage[colorlinks=true,linkcolor=blue,citecolor=blue,urlcolor=blue]{hyperref}

\KOMAoptions{
	headings=standardclasses,%
    toc=listof,%
	toc=bib,%
    numbers=noenddot,%
}

\addtokomafont{captionlabel}{\bfseries}
\setcapwidth[c]{1.0\textwidth}

\KOMAoptions{headsepline=true}
\setkomafont{pageheadfoot}{\textsc}
\ihead{Phase-Field Peridynamics}
\ohead{{\hypersetup{linkcolor=black} \thepage\,/\,\pageref{LastPage}}}
\ofoot[{\hypersetup{linkcolor=black} \thepage\,/\,\pageref{LastPage}}]{}
\numberwithin{equation}{section}

\addto\extrasenglish{%
}

\makeatletter
\renewcommand*\@pnumwidth{2.2em}
\makeatother

\definecolor{kblue}{RGB}{0,158,212}
\definecolor{commentgreen}{RGB}{0,128,0}
\newcommand*{\equationref}[1]{\hyperref[#1]{Eq.~(\ref*{#1})}}
\newcommand*{\appref}[1]{\hyperref[#1]{Appendix~\ref*{#1}}}
\newlength{\textwidthdissertation}
\newcommand{\figsizefactor}{0.917} 
\newcommand{\figsize}[1]{\dimexpr\figsizefactor\dimexpr#1\relax\relax}

\newcommand*{\vb}{\bm}                      
\newcommand*{\abs}[1]{\left|#1\right|}     

\newcommand*{\eqperiod}{\; .}               
\newcommand*{\eqcomma}{\; ,}                

\newcommand*{\transpose}{^{\top}}           

\newcommand*{\identitytensor}{\vb{I}}       
\newcommand*{\zerotensor}{\vb{0}}           

\newcommand*{\dd}{\mathop{}\!\mathrm{d}}    
\newcommand*{\dv}[2]{\frac{\mathrm{d}#1}{\mathrm{d}#2}}           
\newcommand*{\pdv}[2]{\frac{\partial#1}{\partial#2}}               

\newcommand*{\gradient}{\mathop{}\!\nabla}              
\newcommand*{\divergence}{\mathop{}\!\gradient\cdot}    

\newcommand*{\body}{\mathcal{B}}            
\newcommand*{\bodyd}{\mathcal{B}^N}         

\newcommand*{\neumannboundary}{\partial_{\mathrm{N}}\body_0}     
\newcommand*{\dirichletboundary}{\partial_{\mathrm{D}}\body_0}   
\newcommand*{\bextN}{\vb{\bar{B}}_{\mathrm{N}}}                  

\newcommand*{\horizon}{\delta}              
\newcommand*{\horizonratio}{m_{\horizon}}   
\newcommand*{\family}{\mathcal{H}}          
\newcommand*{\familyd}{\mathcal{H}^N}       
\newcommand*{\familydk}{\mathcal{H}^N_\didx}   

\newcommand*{\familyvolume}{V^{\mathcal{H}}}

\newcommand*{\azimuthalangle}{\phi}         
\newcommand*{\polarangle}{\varphi}          

\newcommand*{\didx}{k}                      
\newcommand*{\didxn}{n}                     

\newcommand*{\Vd}{V_\didx}                  
\newcommand*{\Vneighbord}{V_\didxn}         
\newcommand*{\ddVneighbor}{\dd{V'}}         

\newcommand*{\pointspacing}{{\vartriangle} x}	

\newcommand*{\X}{\vb{X}}                    
\newcommand*{\Xd}{\vb{X}_\didx}             
\newcommand*{\Xneighbor}{\X'}               
\newcommand*{\Xneighbord}{\X_\didxn}        

\newcommand*{\x}{\vb{x}}                    
\newcommand*{\xd}{\vb{x}_\didx}             
\newcommand*{\xneighbor}{\x'}               

\newcommand*{\Displacement}{\vb{U}}         

\newcommand*{\Bond}{\vb{\Delta X}}          
\newcommand*{\scalarBond}{\Delta X}          
\newcommand*{\Bondd}{\Bond_{\didx\didxn}}   
\newcommand*{\Bondl}{\Bond}                 
\newcommand*{\bond}{\vb{\Delta x}}          
\newcommand*{\bondd}{\bond_{\didx\didxn}}   

\newcommand*{\DispBond}{\vb{\Delta U}}      
\newcommand*{\DispBondd}{\DispBond_{\didx\didxn}}      

\newcommand*{\Velocity}{\dot{\vb{U}}}       
\newcommand*{\Acceleration}{\ddot{\vb{U}}}  

\newcommand*{\pddefgrad}{\vb{\bar{F}}}                        
\newcommand*{\pdbonddefgrad}{\vb{\tilde{F}}}                  
\newcommand*{\pdbonddefgradavg}{\pdbonddefgrad^{\mathrm{avg}}}  

\newcommand*{\materialbodyforce}{\vb{\bar{B}}}  
\newcommand*{\bint}{\materialbodyforce_{\mathrm{int}}}        
\newcommand*{\bext}{\materialbodyforce_{\mathrm{ext}}}        

\newcommand*{\firstpiola}{\vb{P}}            
\newcommand*{\pdbondfirstpiola}{\vb{\tilde{P}}}          

\newcommand*{\forcestate}{\vb{t}}            

\newcommand*{\shapetensor}{\vb{K}}           
\newcommand*{\bondshapevector}{\vb{B}}     

\newcommand*{\totalenergy}{\mathcal{E}}      

\newcommand*{\strainenergydensity}{\Psi}     
\newcommand*{\pdstrainenergydensity}{\bar{\strainenergydensity}}      

\newcommand*{\density}{\varrho}              
\newcommand*{\densityref}{\density_0}        
\newcommand*{\youngsmodulus}{E}              
\newcommand*{\poissonsratio}{\nu}            

\newcommand*{\bondstrain}{\varepsilon_b}     

\newcommand*{\kernel}{\omega}                
\newcommand*{\kernelscalar}{\omega_s}        

\newcommand*{\rkmonomialvec}{\vb{Q}}         
\newcommand*{\rkmonomialvecn}{\rkmonomialvec_{[n]}}        
\newcommand*{\rkmonomialveczero}{\rkmonomialvec^{\nabla}}  
\newcommand*{\rkmonomialveczeron}{\rkmonomialveczero_{[n]}} 
\newcommand*{\rkmomentmatrix}{\vb{M}}        
\newcommand*{\rkmomentmatrixn}{\vb{M}_{[n]}} 

\newcommand*{\rkshapefunction}{\phi}

\newcommand*{\rkshapefunctionn}{\rkshapefunction_{\didxn}}

\newcommand*{\pointdamage}{D}
\newcommand*{\pointdamaged}{D_\didx}

\newcommand*{\bondfailured}{d_{\didx\didxn}}

\newcommand*{\criticalbondstrain}{\varepsilon_c}

\newcommand*{\baqpnonunistresstensor}{\vb{Z}} 
\newcommand*{\avgbondkernel}{\bar{\omega}_b} 

\newcommand*{\classicalphasefield}{s}
\newcommand*{\phasefield}{s}
\newcommand*{\phasefieldd}{s_{\didx\didxn}} 
\newcommand*{\phasefieldpoint}{\bar{s}}

\newcommand*{\phasefieldcrit}{s_c}
\newcommand*{\griffith}{G_c}
\newcommand*{\cracksurface}{\Gamma_c}
\newcommand*{\degradation}{g}
\newcommand*{\kindegradation}{h}
\newcommand*{\lengthscaleparam}{\ell_c}
\newcommand*{\normalizationconstant}{c_0}
\newcommand*{\cracksurfenergyfunction}{\gamma_c}
\newcommand*{\undamagedstrainenergydensity}{\strainenergydensity_0}

\newcommand*{\tensilestrainenergydensity}{\strainenergydensity_0^+}
\newcommand*{\undamagedfirstpiola}{\firstpiola_0}
\newcommand*{\crackdrivingforce}{Y}
\newcommand*{\crackdrivingforcedk}{Y_{\didx\didxn}}

\newcommand*{\criticalcrackdrivingforce}{Y_c}
\newcommand*{\historyvariable}{\mathcal{Y}}
\newcommand*{\historyvariabled}{\mathcal{Y}_{\didx\didxn}}
\newcommand*{\totalcrackdissipation}{\mathcal{E}^{\Gamma}}
\newcommand*{\crackdissipationpoint}{E_{\X}^\Gamma}
\newcommand*{\crackdissipationpointatz}{E_{z}^\Gamma}
\newcommand*{\crackdissipationbond}{E_{b}^\Gamma}
\newcommand*{\mechenergybond}{E_b}
\newcommand*{\crackarea}{A_\Gamma}
\newcommand*{\cracknormalvector}{\vb{n}_\Gamma}
\newcommand*{\crackcoordinate}{z}
\newcommand*{\cracknormcoordinate}{\xi}
\newcommand*{\sphericalcap}{\mathcal{C}}
\newcommand*{\sphericalcapatz}{\mathcal{C}_{z}}
\newcommand*{\capvolume}{V^{\sphericalcap}}
\newcommand*{\volfracfunction}{f_V}

\newcommand*{\kernelcapfrac}{f_\omega}
\newcommand*{\kernelintegral}{\omega_0}

\AtEveryBibitem{%
  \clearname{translator}%
  \clearlist{publisher}%
  \clearfield{pagetotal}%
}
\renewbibmacro{in:}{}
\AtBeginBibliography{%
  \emergencystretch=1em 
}
\theoremstyle{definition}

\SetKwComment{jlcomment}{\# }{}

\SetCommentSty{mycommfont}

\newcommand*{\edit}[1]{} 

\newcommand*{\afterdefense}[1]{}

\title{\vspace*{-15mm}Phase-Field Peridynamics}
\author{%
Kai Partmann$^{1}$\footnote{Corresponding author: kai.partmann@uni-siegen.de},
Christian Wieners$^2$,
Michael Ortiz$^3$,
and Kerstin Weinberg$^1$
}
\date{%
\normalsize
$^1$ University of Siegen, Germany\\
$^2$ Karlsruhe Institute of Technology (KIT), Germany\\
$^3$ California Institute of Technology (Caltech), USA\\[\bigskipamount]
\today
}

\begin{document}

\maketitle

\begin{addmargin}{12pt}
\begin{abstract}
\noindent
Peridynamics formulates the balance of linear momentum as an integro-differential equation, making it naturally suited for fracture modeling without special treatment of discontinuities.
The bond-associated correspondence formulation provides a highly accurate peridynamic framework by computing bond-wise deformation gradients that are free of zero-energy modes and yield accurate results even near boundaries.
However, the traditional fracture approach based on irreversible bond deletion can compromise this formulation, as the progressive removal of bonds degrades the nonlocal approximation of the deformation gradient and can lead to numerical instabilities.
In this work, a novel phase-field peridynamics approach is introduced that avoids these instabilities.
Instead of deleting bonds, the energetic contribution of each bond is continuously degraded through a bond phase-field parameter, while a separate kinematic degradation function preserves the accuracy of the nonlocal deformation gradient approximation.
The normalization constant ensuring thermodynamic consistency with Griffith's fracture theory is derived analytically for general spherical kernel functions as a ratio of two one-dimensional integrals.
Numerical examples including mode~I and mode~II fracture, the boundary tension test with different kernel functions and horizon ratios, and the Kalthoff-Winkler experiment demonstrate the stability, accuracy, and consistency of the proposed approach.
\end{abstract}
\end{addmargin}

\bigskip


\section{Introduction}

Predicting how cracks initiate and propagate through load-bearing structures is one of the central challenges in computational mechanics.
Whether fracture is to be prevented in safety-critical applications or deliberately induced in manufacturing processes, its accurate computational prediction at realistic loading rates and deformation levels is of fundamental importance.
The foundations for fracture prediction were established by the pioneering work of Griffith~\cite{Griffith1921}, who proposed that a crack grows when the energy released by the advancing crack exceeds the energy required to create new crack surfaces.
Irwin~\cite{Irwin1958} extended this framework by introducing the stress intensity factor, making fracture mechanics applicable to a broader class of materials.
These energy-based approaches remain at the core of modern fracture theory.

Numerous computational approaches have been developed to address the challenges of fracture modeling, each with distinct strengths and limitations.
In the standard finite element method (FEM), cracks represent discontinuities in the displacement field that must be explicitly resolved by the mesh, requiring costly remeshing procedures as cracks propagate~\cite{Belytschko1985,Bonet2008}.
The extended finite element method (XFEM)~\cite{BelytschkoBlack1999,Moes1999} alleviates this by enriching the approximation with discontinuous functions, though it becomes increasingly complex for crack branching and three-dimensional problems.
Cohesive zone models~\cite{Dugdale1960,Barenblatt1962,CamachoOrtiz1996,OrtizPandolfi1999} describe fracture through traction-separation laws governing the progressive decohesion of material surfaces, but distributing cohesive elements throughout the mesh introduces mesh sensitivity in the crack trajectory.
The phase-field approach to fracture~\cite{Francfort1998,Ambrosio1990,Bourdin2000,Miehe2010} regularizes the sharp crack topology with a continuous damage field, enabling the natural modeling of crack initiation, propagation, and branching without explicit crack tracking.
Extensions to dynamic fracture have been developed by Borden et al.~\cite{Borden2012} and Weinberg and Wieners~\cite{Weinberg2022}, including approaches combining phase-field methods with peridynamics~\cite{FriebertshaeuserPartmann2023,FriebertshaeuserPartmann2023a}.
However, the phase-field method requires a fine mesh resolution around the crack to resolve the diffuse damage band, particularly in three-dimensional dynamic settings.
Further comparisons of computational fracture approaches have been provided by Rabczuk~\cite{Rabczuk2013}, Pandolfi et al.~\cite{PandolfiWeinbergOrtiz2021}, and Diehl et al.~\cite{Diehl2022}.

All of the above approaches are based on classical continuum mechanics, where the governing equations are formulated as partial differential equations that require spatial derivatives of the displacement field.
At a crack, the displacement field is discontinuous and these derivatives are undefined, therefore all classical methods require special techniques to handle the presence of cracks.
Peridynamics, introduced by Silling~\cite{Silling2000}, takes a fundamentally different approach by formulating the balance of linear momentum as an integro-differential equation.
Internal forces at a material point are computed by integrating the interactions with all neighboring points within a finite distance, called the horizon.
While the spherical neighborhood is the standard choice, alternative neighborhood shapes such as cube-shaped horizons have also been explored~\cite{Partmann2026}.
Since no spatial derivatives of the displacement field are required, the governing equations remain valid at discontinuities such as cracks, making peridynamics naturally suited for fracture modeling~\cite{Silling2005,Madenci2014}.
A rigorous variational theory for integral functionals based on finite-horizon nonlocal gradients, including $\Gamma$-convergence results that recover classical local models, has been developed by Cueto et al.~\cite{Cueto2023,Cueto2025}.
Energy-based peridynamic fracture models with continuous damage that converge to Griffith fracture in the vanishing-horizon limit have been proposed by Lipton~\cite{Lipton2014,Lipton2015,Lipton2024}.

The original bond-based (BB) formulation~\cite{Silling2000} is simple and robust but restricts the Poisson's ratio.
The ordinary state-based (OSB) formulation~\cite{Silling2007} overcomes this restriction but still suffers from the surface effect, a systematic error near boundaries caused by the truncation of the neighborhood.
The correspondence formulation~\cite{Silling2007} approximates the deformation gradient through nonlocal weighted averaging, enabling the direct use of classical constitutive models and eliminating the surface effect.
However, it introduces stability issues due to zero-energy modes~\cite{Silling2017,Breitenfeld2014}.
The bond-associated approach~\cite{Chen2018,Chen2019,Behzadinasab2021,Partmann2025} addresses this instability by computing bond-wise deformation gradients that are inherently free of zero-energy modes.
This makes the bond-associated correspondence formulation the most accurate peridynamic formulation that consistently reproduces the elastic behavior of the classical continuum.

Fracture in peridynamics is traditionally modeled by irreversible bond deletion, where a bond is permanently removed when its deformation exceeds a critical threshold derived from Griffith's energy criterion~\cite{Silling2005}.
This approach has been successfully used to simulate dynamic crack propagation and branching with the standard formulations~\cite{Ha2010,Silling2002,FriebertshaeuserPartmann2022,FriebertshaeuserPartmann2022a,Khosravani2022}.
However, in correspondence-based formulations, the progressive deletion of bonds degrades the shape tensor required for approximating the deformation gradient.
As more bonds are removed in the vicinity of a propagating crack, the shape tensor can become ill-conditioned or singular, potentially causing a breakdown of the kinematic approximation and numerical instabilities.
These stability issues can limit the reliable application of the most accurate peridynamic formulation to fracture problems.

In this work, a novel phase-field peridynamics (PFPD) approach is introduced that avoids these stability issues.
Instead of deleting bonds instantaneously when a critical deformation is reached, the energetic contribution of each bond is degraded continuously through a bond phase-field parameter.
The damage evolution follows from an energy-based variational principle and ensures thermodynamic consistency with Griffith's fracture theory.
A separate kinematic degradation function is introduced to maintain the accuracy of the shape tensor even under significant damage, preventing the numerical instabilities that can arise from bond deletion.
The normalization of the damage model is derived analytically for general spherical kernel functions by requiring that the total crack dissipation equals the Griffith energy release rate per unit crack area.

The paper is organized as follows.
\autoref{sec:peridynamics} provides a brief introduction to the bond-associated correspondence formulation.
In \autoref{sec:pfpd}, the phase-field peridynamics framework is developed, including the bond-associated phase-field approximation, the damage evolution law, and the analytical derivation of the normalization constant.
\autoref{sec:numerics} presents the discretization and algorithmic implementation, with a detailed comparison to the standard bond-associated approach.
Numerical examples are presented in \autoref{sec:examples}, including mode~I and mode~II fracture, the verification of the normalization constant, and the Kalthoff-Winkler experiment.
The paper concludes with a summary and discussion of the results in \autoref{sec:conclusion}.

\section{Peridynamics Theory}\label{sec:peridynamics}

This section provides a brief introduction to the peridynamic framework in a continuous setting, focusing on the bond-associated quadrature point (BAQP) correspondence formulation that forms the basis for the phase-field peridynamics model developed in the subsequent sections.

\subsection{Nonlocal kinematics and balance of linear momentum}

\begin{figure}[ht]
\centering
\includegraphics[width=\figsize{0.7\textwidthdissertation}]{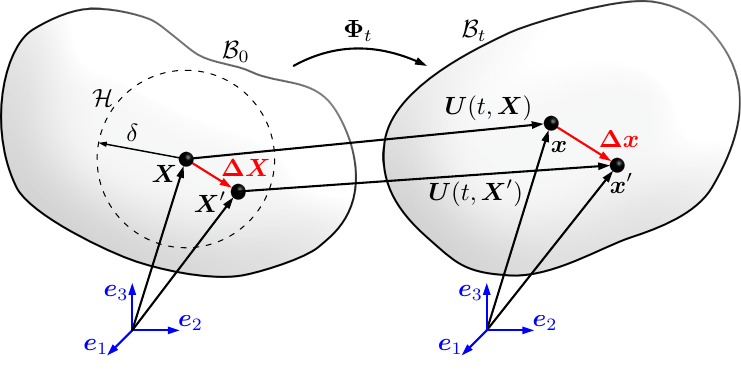}
\caption{Deformation of a peridynamic body $\body$ from reference configuration $\body_0$ to current configuration $\body_t$. The neighborhood~$\family(\X)$ defines the nonlocal interaction domain around each material point~$\X$.}
\label{fig:pd_deformation}
\end{figure}

Consider a body~$\body$ occupying the reference configuration~$\body_0 \subset \mathbb{R}^3$ and the current configuration~$\body_t \subset \mathbb{R}^3$.
In peridynamics, the body is represented as a continuum of material points $\body_0\subset\mathbb{R}^3$, interacting with each other through nonlocal forces~\cite{Silling2000}.
The interaction between a point~$\X$ and its neighbor~$\Xneighbor$ is characterized by a \emph{bond}~$\Bond = \Xneighbor - \X$ in the reference configuration and by $\bond = \xneighbor - \x$ in the current configuration.
Each point~$\X$ interacts only with points inside a specified \emph{neighborhood} or \emph{point family}
\begin{equation}\label{eq:family}
\family(\X) = \left\{\Xneighbor \in \body_0 \; \big| \;  0 < \abs{\Bondl} \leq \horizon \right\} \eqcomma
\end{equation}
defined as all points within the \emph{horizon}~$\horizon \in \mathbb{R}^+$.
The \emph{bond displacement} is defined as $\DispBond = \Displacement(t,\Xneighbor) - \Displacement(t,\X)$, with the relation $\bond = \Bond + \DispBond$.
The peridynamic kinematics are illustrated in \autoref{fig:pd_deformation}.

The balance of linear momentum in peridynamics reads
\begin{equation}\label{eq:balancelinearmomentum}
\densityref(\X) \, \Acceleration(t,\X) = \bint(t,\X) + \bext(t,\X) \qquad \forall (t,\X) \in [0,\infty) \times \body_0 \eqcomma
\end{equation}
where $\bint$ denotes the internal body force density and $\bext$ the external body force density.
The initial conditions $\Displacement(0,\X) = \Displacement_0(\X)$ and $\Velocity(0,\X) = \Velocity_0(\X)$ specify the displacement and velocity at $t=0$.
Boundary conditions are imposed by prescribing either displacement (Dirichlet) or external force density (Neumann) on a boundary layer of material points with finite thickness, since the integro-differential equation~\equationref{eq:balancelinearmomentum} requires a volumetric treatment rather than classical surface boundary conditions~\cite{Silling2000}.

The internal force density approximates the divergence of the first Piola-Kirchhoff stress tensor with
\begin{equation}
\bint(t,\X) = \divergence{\firstpiola}(t,\X) + \mathcal{O}(\horizon^2) \eqperiod
\end{equation}
Since the governing equation \equationref{eq:balancelinearmomentum} is an integro-differential equation that does not involve spatial derivatives of the displacement field, it remains valid at discontinuities such as cracks, making peridynamics naturally suited for fracture modeling.

\subsection{The correspondence formulation}

The correspondence formulation~\cite{Silling2007} approximates the deformation gradient through nonlocal weighted averaging over the neighborhood, enabling the direct use of classical constitutive models.
An approximated deformation gradient~$\pddefgrad$ is defined as
\begin{equation}\label{eq:pddefgrad}
\pddefgrad(t,\X) = \left( \int_{\family(\X)} \kernel(\Bondl) \, \bond \otimes \Bond \ddVneighbor \right) \shapetensor^{-1}(\X) \eqcomma
\end{equation}
where $\kernel \colon \{\mathbb{R}^3 \setminus {\vb{0}}\} \rightarrow [0,\infty) $ with $\kernel(\Bond) = 0$ for $\abs\Bond > \horizon$ is a kernel (or influence) function and the \emph{shape tensor}~$\shapetensor$ is defined as
\begin{equation}\label{eq:shapetensor}
\shapetensor(\X) = \int_{\family(\X)} \kernel(\Bondl) \, \Bondl \otimes \Bondl \ddVneighbor \eqperiod
\end{equation}
%
The shape tensor is a symmetric and positive definite matrix for bulk points with full neighborhoods.
Due to the multiplication with~$\shapetensor^{-1}$, the deformation gradient is exactly reproduced for all homogeneous deformations, independent of the neighborhood geometry.
This means that the correspondence formulation does not suffer from the surface effect, a shortcoming of the standard bond-based and ordinary state-based formulations~\cite{Le2018,Madenci2014}.

With the deformation gradient, the stress tensor of each point can be evaluated from classical constitutive laws.
A \emph{bond shape vector}~$\bondshapevector \colon \body_0 \times \body_0 \to \mathbb{R}^3$ is defined as
\begin{equation}\label{eq:bondshapevector}
\bondshapevector(\X, \Xneighbor) = \kernel(\Bondl) \, \shapetensor^{-1}(\X) \, \Bond \eqperiod
\end{equation}
With this, the deformation gradient and the \emph{force vector state}~$\forcestate$ can be compactly expressed as
\begin{equation}\label{eq:pddefgradbondshapevector}
\pddefgrad(t,\X) = \int_{\family(\X)} \bond \otimes \bondshapevector(\X, \Xneighbor) \ddVneighbor \eqcomma
\end{equation}
and, depending on the  first Piola-Kirchhoff stress tensor $\firstpiola(t,\X)= \firstpiola(\pddefgrad(t,\X))$,
\begin{equation}\label{eq:forcestate}
\forcestate(t,\X,\Xneighbor) = \firstpiola(t,\X) \, \bondshapevector(\X, \Xneighbor) \eqperiod
\end{equation}
The divergence of the first Piola-Kirchhoff stress tensor is then approximated as
\begin{equation}\label{eq:pdinternalforcedensity}
\bint(t,\X) = \int_{\family(\X)} \big( \forcestate(t,\X,\Xneighbor) - \forcestate(t,\Xneighbor,\X) \big) \ddVneighbor \eqperiod
\end{equation}
While the correspondence formulation overcomes the limitations of the bond-based and ordinary state-based models, it introduces stability issues from zero-energy modes~\cite{Silling2017,Breitenfeld2014}, which are oscillatory displacement patterns that produce a vanishing deformation gradient and therefore vanishing internal forces~\cite{Littlewood2010}.
The bond-associated approach addresses this shortcoming by computing bond-wise deformation gradients that capture local deformation variations.

\subsection{Bond-associated quadrature points}\label{sec:baqp}

The bond-associated quadrature point (BAQP) approach, introduced by Behzadinasab et al.~\cite{Behzadinasab2021,Behzadinasab2021a,Bazilevs2022}, addresses the zero-energy mode instability by computing deformation gradients at bond-associated quadrature points located at the center of each bond.
This is conceptually similar to higher-order quadrature in the finite element method, where additional integration points suppress hourglass modes.

A \emph{bond-associated deformation gradient}~$\pdbonddefgrad$ is defined for each bond as
\begin{equation}\label{eq:baqpbonddefgrad}
\pdbonddefgrad(t,\X,\Xneighbor) = \pdbonddefgradavg(t,\X,\Xneighbor) + \left(\bond - \pdbonddefgradavg(t,\X,\Xneighbor) \, \Bond\right) \otimes \frac{\Bond}{\abs{\Bond}^2} \eqcomma
\end{equation}
where the average deformation gradient of a bond is
\begin{equation}
\pdbonddefgradavg(t,\X,\Xneighbor) = \frac{1}{2} \left(\pddefgrad(t,\X) + \pddefgrad(t,\Xneighbor)\right) \eqperiod
\end{equation}
The correction term in \equationref{eq:baqpbonddefgrad} ensures that the bond deformation gradient exactly reproduces the actual bond deformation in the bond direction, while the average provides the transverse deformation components.
This construction eliminates zero-energy modes, since oscillatory displacement patterns that would cancel in the point-wise average are captured by the bond-wise correction~\cite{Behzadinasab2021}.

The strain energy density of a point is approximated by averaging the bond-wise strain energy densities over the neighborhood with
\begin{equation}\label{eq:baqpstrainenergydensity}
\pdstrainenergydensity(t,\X) = \int_{\family(\X)} \avgbondkernel(\X,\Xneighbor) \, \strainenergydensity(\pdbonddefgrad(t,\X,\Xneighbor)) \ddVneighbor \eqcomma
\end{equation}
where the \emph{averaged bond kernel function}~$\avgbondkernel \colon \body_0 \times \body_0 \to \mathbb{R}^+$ is defined as
\begin{equation}\label{eq:baqpavgbondkernel}
\avgbondkernel(\X,\Xneighbor) = \frac{1}{2} \left(\frac{\kernel(\Bond)}{\kernelintegral(\X)} + \frac{\kernel(-\Bond)}{\kernelintegral(\Xneighbor)}\right) \eqcomma
\end{equation}
with the \emph{kernel integral}~$\kernelintegral \colon \body_0 \rightarrow \mathbb{R}^+$ defined as
\begin{equation}\label{eq:kernelintegraldef}
\kernelintegral(\X) = \int_{\family(\X)} \kernel(\Bond) \ddVneighbor \eqperiod
\end{equation}
The kernel integral satisfies the normalization condition for all interior points as
\begin{equation}\label{eq:kernelintegralnormalization}
\kernelintegral(\X) = 1 \qquad \forall \X \in \left\{ \X \in \body_0 \;\middle|\; \family(\X) \subset \body_0 \right\} \eqperiod
\end{equation}
This normalization condition is just a statement about internal points, since $\kernelintegral$ appears in the denominator of the averaged bond kernel function~\eqref{eq:baqpavgbondkernel} so boundary truncation is correctly accounted for.
A commonly used kernel~\cite{Behzadinasab2021} satisfying this condition is the cubic spline kernel
\begin{equation}\label{eq:baqpcubicsplinekernel}
\kernel(\Bond) = \frac{8}{\pi \horizon^3} \cdot \begin{cases}
1 - 6\left(\frac{\abs{{\Bond}}}{\horizon}\right)^2 + 6\left(\frac{\abs{{\Bond}}}{\horizon}\right)^3 &\quad \text{if} \ \; \frac{\abs{{\Bond}}}{\horizon} \leq \frac{1}{2} \eqcomma\\[2mm]
2\left(1 - \frac{\abs{{\Bond}}}{\horizon}\right)^3 &\quad \text{if} \ \; \frac{1}{2} < \frac{\abs{{\Bond}}}{\horizon} \leq 1 \eqcomma\\[2mm]
0 &\quad \text{if} \ \; \frac{\abs{{\Bond}}}{\horizon} > 1 \eqperiod
\end{cases}
\end{equation}

The bond-wise first Piola-Kirchhoff stress tensor is evaluated from the constitutive model as
\begin{equation}\label{eq:baqppdbondfirstpiola}
\pdbondfirstpiola(t,\X,\Xneighbor) = \firstpiola(\pdbonddefgrad(t,\X,\Xneighbor)) \eqperiod
\end{equation}
Introducing the \emph{non-uniform stress tensor}~$\baqpnonunistresstensor \colon [0,\infty) \times \body_0 \rightarrow \mathbb{R}^{3\times3}$ as
\begin{equation}\label{eq:baqpnonunistresstensor}
\baqpnonunistresstensor(t,\X) = \int_{\family(\X)} \avgbondkernel(\X,\Xneighbor) \, \pdbondfirstpiola(t,\X,\Xneighbor) \left( \identitytensor - \frac{\Bond \otimes \Bond}{\abs{\Bond}^2} \right) \ddVneighbor \eqcomma
\end{equation}
the force state can be written compactly as
\begin{equation}\label{eq:baqpforcestate}
\forcestate(t,\X,\Xneighbor) = \avgbondkernel(\X,\Xneighbor) \, \pdbondfirstpiola(t,\X,\Xneighbor) \, \frac{\Bond}{\abs{\Bond}}
+ \baqpnonunistresstensor(t,\X) \, \bondshapevector(\X,\Xneighbor) \eqperiod
\end{equation}
The first term projects the bond stress along the bond direction, while the second term distributes the non-uniform stress correction via the bond shape vector.
This formulation is inherently free of zero-energy modes and provides a stable correspondence framework for large deformation problems~\cite{Behzadinasab2021,Bazilevs2022}.

\section{Phase-Field Peridynamics}\label{sec:pfpd}

Following the brittle fracture approaches of Griffith and Irwin \cite{Griffith1921,Irwin1958}, a material fails when a critical energy release rate is reached.
This assumption provides a criterion for crack growth based on energy considerations, which has to be embedded into an energy minimization setting for the entire structure to determine crack initiation and propagation \cite{Miehe2010,Weinberg2022,Partmann2023,FriebertshaeuserPartmann2023}.
Consequently, the total energy of a body is composed of its kinetic energy, the potential energy with the strain energy density of the undamaged material~$\undamagedstrainenergydensity$, and the surface energy contributions from evolving crack boundaries with the critical energy release rate~$\griffith$.
Here, $\undamagedstrainenergydensity$ denotes the strain energy density~$\strainenergydensity$ from \autoref{sec:peridynamics}, now carrying a subscript~$0$ to distinguish the undamaged material response from the degraded one in the phase-field setting.
With this assumption, the total energy functional reads
\begin{equation}
\totalenergy(t) = \int_{\body_0 \setminus \cracksurface(t)} \left(\frac{1}{2} \densityref \abs{\Velocity}^2 + \undamagedstrainenergydensity \right) \dd{V} + \int_{\cracksurface(t)} \griffith \; \dd{A} \eqcomma
\end{equation}
where $\cracksurface(t)$ is the crack surface at time $t$.
An important assumption for thermodynamic consistency is that the crack evolution is irreversible, i.e., $\cracksurface(t_1) \subset \cracksurface(t_2)$ for $t_1 < t_2$.

In the \textbf{classical phase-field approach} \cite{Ambrosio1990,Bourdin2000,Miehe2010}, the sharp crack topology is regularized via a diffusive crack approximation.
To characterize the state of the material, a phase-field parameter~$\classicalphasefield$ is introduced.
For the approximation, the discontinuous set of evolving crack surfaces~$\cracksurface(t)$ is replaced by a continuous surface energy density function, approximating the surface integral via a volumetric integral over the entire body~$\body_0$.
This leads to the objective functional for the total energy as
\begin{align}
\totalenergy(t) &= \int_{\body_0} \left(\frac{1}{2} \densityref \abs{\Velocity}^2 + \degradation(\classicalphasefield) \, \undamagedstrainenergydensity + \frac{\griffith}{\normalizationconstant} \left(\frac{\cracksurfenergyfunction(\classicalphasefield)}{\lengthscaleparam} + \lengthscaleparam \abs{\gradient \classicalphasefield}^2 \right)\right) \dd{V} \label{eq:classicalphasefieldtotalenergy}\\
&\rightarrow \text{stationary} \eqperiod \nonumber
\end{align}
The \emph{degradation function}~$\degradation$ quantifies the degradation of the material response due to damage.
Also, the \emph{crack surface density function}~$\cracksurfenergyfunction$ approximates the crack surface energy contribution based on the phase-field parameter.
The regularization term with the \emph{length scale parameter}~$\lengthscaleparam$ ensures a diffusive crack approximation over a finite width.
The \emph{normalization constant}~$\normalizationconstant\in\mathbb{R}^+$ ensures that the total crack dissipation corresponds to Griffith's assumption of energy release rate per unit crack area and can be calculated analytically for a given choice of $\cracksurfenergyfunction$, as known from the literature \cite{Pham2010,Li2023}.
For the commonly used AT1 model with $\cracksurfenergyfunction(\phasefield) = \phasefield$, the normalization constant is $\normalizationconstant = \frac{8}{3}$, while for the AT2 model with $\cracksurfenergyfunction(\phasefield) = \phasefield^2$, one obtains $\normalizationconstant = 2$~\cite{Miehe2010,Pham2010,Li2023}.

In the following, the classical phase-field approach to fracture is adapted to the bond-associated correspondence framework described in \autoref{sec:peridynamics}.
The length scale parameter~$\lengthscaleparam$ is defined as
\begin{equation}
\lengthscaleparam := \horizon \eqcomma
\end{equation}
to maintain consistency with the nonlocal interaction radius in peridynamics.
Since the nonlocal averaging over the horizon inherently provides a regularization of the crack approximation, the regularization term $\lengthscaleparam \abs{\gradient \classicalphasefield}^2$ vanishes in the bond-associated formulation.

\subsection{A bond-associated phase-field approximation}

To embed the phase-field fracture approach into the bond-associated correspondence framework, the total energy functional is expressed in terms of bond contributions.
The \emph{bond phase-field parameter}~$\phasefield \colon [0,\infty) \times \body_0 \times \body_0 \rightarrow [0,1]$ is introduced as a bond-wise variable, defining the state of damage of the bond between two material points~$\X$ and~$\Xneighbor$ at time~$t$.
Following the definition of the point-wise strain energy of the BAQP formulation in \equationref{eq:baqpstrainenergydensity}, the total energy functional from \equationref{eq:classicalphasefieldtotalenergy} can be expressed in terms of bond contributions in a peridynamic setting as
\begin{align}
\totalenergy(t) &= \int_{\body_0} \frac{1}{2} \densityref \abs{\Velocity(t,\X)}^2 \dd{V}\nonumber\\
&+ \int_{\body_0} \int_{\family(\X)} \avgbondkernel(\X,\Xneighbor) \left(\degradation(\phasefield(t,\X,\Xneighbor)) \, \undamagedstrainenergydensity + \frac{\griffith \, \cracksurfenergyfunction(\phasefield(t,\X,\Xneighbor))}{\normalizationconstant \, \horizon}\right) \ddVneighbor \dd{V} \nonumber\\
&\rightarrow \text{stationary} \eqperiod 
\end{align}
To keep the notation consistent with the classical peridynamics approach, a fully damaged bond has the value of $\phasefield = 1$, while an undamaged bond has the value of $\phasefield = 0$.
The usage of the BAQP framework and the additional integral over the horizon inherently provides a regularization of the crack approximation over the horizon length scale.

The \emph{total crack dissipation}~$\totalcrackdissipation \colon [0,\infty) \rightarrow \mathbb{R}$ accounts for the total energy dissipated due to crack formation and growth in the body with
\begin{align}
\totalcrackdissipation(t) &= \int_{\body_0} \int_{\family(\X)} \avgbondkernel(\X,\Xneighbor) \, \frac{\griffith \, \cracksurfenergyfunction(\phasefield(t,\X,\Xneighbor))}{\normalizationconstant \, \horizon} \ddVneighbor \dd{V}\nonumber\\
&= \frac{\griffith}{\normalizationconstant \, \horizon} \int_{\body_0} \int_{\family(\X)} \avgbondkernel(\X,\Xneighbor) \, \cracksurfenergyfunction(\phasefield(t,\X,\Xneighbor)) \ddVneighbor \dd{V} \eqcomma
\end{align}
where $\griffith$, $\normalizationconstant$, and $\horizon$ are constant for each point in the body.
A \emph{bond crack dissipation}~$\crackdissipationbond \colon [0,\infty) \times \body_0 \times \body_0 \rightarrow \mathbb{R}$ can be introduced as
\begin{equation}\label{eq:bondcrackdissipation}
\crackdissipationbond(t,\X,\Xneighbor) = \frac{\griffith \, \cracksurfenergyfunction(\phasefield(t,\X,\Xneighbor))}{\normalizationconstant \, \horizon} \eqcomma
\end{equation}
denoting the energetic contribution of a single bond to the total crack dissipation.
With \equationref{eq:bondcrackdissipation}, the total crack dissipation can be expressed as the integral of the bond crack dissipation over the body and its horizon as
\begin{equation}
\totalcrackdissipation(t) = \int_{\body_0} \int_{\family(\X)} \avgbondkernel(\X,\Xneighbor) \, \crackdissipationbond(t,\X,\Xneighbor) \ddVneighbor \dd{V} \eqperiod
\end{equation}
Here, $\crackdissipationbond$ is bond-wise local and does not depend on the neighboring bonds.
For the remainder of this work, the crack surface density function~$\cracksurfenergyfunction$ is chosen as
\begin{equation}\label{eq:at1pfpd}
\cracksurfenergyfunction(\phasefield) := \phasefield(t,\X,\Xneighbor) \eqcomma
\end{equation}
consistent with the AT1 model in the literature \cite{Ambrosio1990,Pham2010,Li2023}.
This leads to a linear increase of the crack surface energy with increasing damage.

The degradation of the material response due to damage is modeled via the \emph{bond degradation function}~$\degradation \colon [0,1] \rightarrow [0,1]$.
A common choice for the degradation function is the quadratic form
\begin{equation}\label{eq:pfpddegradation}
\degradation(\phasefield) = (1 - \phasefield(t,\X,\Xneighbor))^2 \eqcomma
\end{equation}
leading to a smooth degradation of the material response with increasing damage.
The bond-wise first Piola-Kirchhoff stress tensor~$\pdbondfirstpiola$ from \equationref{eq:baqppdbondfirstpiola} is modified to account for the degradation due to damage as
\begin{equation}
\pdbondfirstpiola(t,\X,\Xneighbor) = \degradation(\phasefield(t,\X,\Xneighbor)) \, \undamagedfirstpiola(\pdbonddefgrad(t,\X,\Xneighbor)) \eqcomma
\end{equation}
with the undamaged first Piola-Kirchhoff stress tensor~$\undamagedfirstpiola$, here denoted with a subscript~$0$ in contrast to the previous section.
This ensures that the force density contribution of a bond vanishes smoothly as the bond becomes fully damaged with $\phasefield \rightarrow 1$.

\begin{figure}[ht]
\centering
\includegraphics[width=\figsize{0.7\textwidthdissertation}]{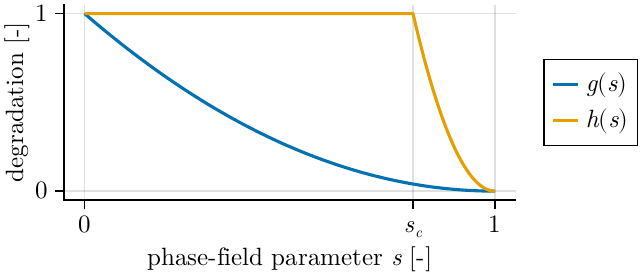}
\caption[The bond and kinematic degradation functions]{The bond and kinematic degradation functions~$\degradation$ and~$\kindegradation$ over the bond phase-field parameter~$\phasefield$.}
\label{fig:pfpd_degradation}
\end{figure}

However, in peridynamics, the kinematic approximation via the shape tensor~$\shapetensor$ and the associated bond shape vectors~$\bondshapevector$ plays a crucial role in ensuring accurate deformation measures that have to be considered when degrading bonds.
To maintain numerical stability and the accuracy of the kinematic approximation, a \emph{bond kinematic degradation function}~$\kindegradation \colon [0,1] \rightarrow [0,1]$ is introduced.
It is defined as
\begin{equation}\label{eq:pfpdkindegradation}
\kindegradation(\phasefield) = \begin{cases}
1 & \text{if }\phasefield(t,\X,\Xneighbor) \leq \phasefieldcrit \eqcomma \\
\left(\frac{1 - \phasefield(t,\X,\Xneighbor)}{1 - \phasefieldcrit}\right)^2 & \text{if }\phasefield(t,\X,\Xneighbor) > \phasefieldcrit \eqcomma
\end{cases}
\end{equation}
where $\phasefieldcrit \in [0,1]$ is a parameter defining a threshold for the degradation of the kinematic approximation, as displayed in \autoref{fig:pfpd_degradation}.
For $\phasefieldcrit = 1$, the kinematic degradation is disabled, while for $\phasefieldcrit = 0$, the kinematic approximation degrades similarly to the energetic contribution with
\begin{equation}
\kindegradation(\phasefield(t,\X,\Xneighbor)) \big|_{\phasefieldcrit = 0} = \degradation(\phasefield(t,\X,\Xneighbor)) \eqperiod
\end{equation}
The shape tensor and bond shape vector are modified to account for the kinematic degradation as
\begin{equation}\label{eq:pfpdshapetensor}
\shapetensor(t,\X) = \int_{\family(\X)} \kernel(\Bondl) \, \kindegradation(\phasefield(t,\X,\Xneighbor)) \, \Bondl \otimes \Bondl \ddVneighbor \eqcomma
\end{equation}
and 
\begin{equation}\label{eq:pfpdbondshapevector}
\bondshapevector(t,\X, \Xneighbor) = \kernel(\Bondl) \, \kindegradation(\phasefield(t,\X,\Xneighbor)) \, \shapetensor^{-1}(\X) \, \Bond \eqperiod
\end{equation}
Note that both the shape tensor and the bond shape vector are now time-dependent due to the time dependence of the phase-field.
Without the kinematic degradation, damage under large deformations can lead to unphysical deformation gradients, and fully broken fragments are incorrectly still connected to the intact material via the kinematic approximation.

\subsection{Damage evolution and crack criteria}

The evolution of the damage state in classical phase-field fracture is derived from a variational principle minimizing the total energy functional with respect to the phase-field parameter~$\classicalphasefield$ \cite{Bourdin2000,Miehe2010}.
Similarly, here a bond-wise local energy minimization principle is employed to derive the evolution of the bond phase-field parameter~$\phasefield$.
Considering a static situation and a single bond between two material points~$\X$ and~$\Xneighbor$, the energy contribution of a bond~$\mechenergybond \colon [0,\infty) \times \body_0 \times \body_0 \rightarrow \mathbb{R}$ is defined as
\begin{equation}
\mechenergybond(t,\X,\Xneighbor) = \degradation(\phasefield(t,\X,\Xneighbor)) \, \undamagedstrainenergydensity(\pdbonddefgrad(t,\X,\Xneighbor)) + \crackdissipationbond(t,\X,\Xneighbor) \eqperiod
\end{equation}
The bond phase-field parameter evolves by minimizing the bond mechanical energy as
\begin{align}
\frac{\delta \mechenergybond}{\delta \phasefield} = \dv{g(s(t,\X,\Xneighbor))}{s} \undamagedstrainenergydensity + \frac{\griffith}{\normalizationconstant \, \horizon} \dv{\cracksurfenergyfunction(\phasefield(t,\X,\Xneighbor))}{\phasefield} = 0 \eqcomma
\end{align}
leading with \equationref{eq:at1pfpd} and \equationref{eq:pfpddegradation} to $\dv{\cracksurfenergyfunction(\phasefield(t,\X,\Xneighbor))}{\phasefield} = 1$ and the relation
\begin{equation}
-2 (1 - \phasefield(t,\X,\Xneighbor)) \, \undamagedstrainenergydensity + \frac{\griffith}{\normalizationconstant \, \horizon} = 0 \eqperiod
\end{equation}
Substituting a phase-field value of $\phasefield(t,\X,\Xneighbor) = 0$ yields the \emph{critical crack driving force}~$\criticalcrackdrivingforce$ as
\begin{equation}\label{eq:criticalcrackdrivingforcestrainenergy}
\criticalcrackdrivingforce := \frac{\griffith}{2 \, \normalizationconstant \, \horizon} \eqcomma
\end{equation}
defining an elastic strain energy density threshold for damage initiation.
Deformations below this threshold do not lead to damage growth, and the bond remains intact with $\phasefield = 0$.
Only the tensile (positive) part of the undamaged strain energy density, denoted~$\tensilestrainenergydensity$, contributes to damage growth, therefore the bond-wise \emph{crack driving force}~$\crackdrivingforce \colon [0,\infty) \times \body_0 \times \body_0 \rightarrow \mathbb{R}$ is defined as
\begin{equation}\label{eq:crackdrivingforcestrainenergy}
\crackdrivingforce(t,\X,\Xneighbor) = \tensilestrainenergydensity(\pdbonddefgrad(t,\X,\Xneighbor)) \eqperiod
\end{equation}
This approach is consistent with well-established phase-field fracture models in the literature \cite{Pham2010,Li2023}.

In the work of Bilgen et al.~\cite{Bilgen2019,Bilgen2019a}, a physically motivated definition of the crack driving force based on the maximum principal stress~$\sigma_1$ is proposed.
The maximum principal stress is computed as the maximum eigenvalue of the bond-wise Cauchy stress tensor.
This leads to the definition of a critical stress~$\sigma_c$ for damage initiation, ensuring that damage only occurs when the maximum principal stress exceeds this critical value.
Defining stress as a driving force for damage is particularly appealing for engineering applications, as design criteria often rely on stress-based failure theories.
The strain energy density of a bond for a linear elastic material is given as
\begin{equation}
\undamagedstrainenergydensity = \frac{\sigma^2}{2 \, \youngsmodulus} \eqcomma
\end{equation}
with the Cauchy stress~$\sigma$ and Young's modulus~$\youngsmodulus$.
The critical stress~$\sigma_c$ is derived from the critical strain energy density in \equationref{eq:criticalcrackdrivingforcestrainenergy} as
\begin{equation}
\sigma_c := \sqrt{\frac{\youngsmodulus \, \griffith}{\normalizationconstant \, \horizon}} \eqperiod
\end{equation}
Analogously, the critical crack driving force is expressed in terms of the critical stress as
\begin{equation}
\criticalcrackdrivingforce := \frac{{\sigma_c}^2}{2 \, \youngsmodulus} \eqperiod
\end{equation}
The crack driving force is then defined based on the maximum principal stress as
\begin{equation}\label{eq:crackdrivingforcestress}
\crackdrivingforce(t,\X,\Xneighbor) = \frac{\big\langle\sigma_1(\pdbonddefgrad(t,\X,\Xneighbor))\big\rangle_+^2}{2 \, \youngsmodulus} \eqcomma
\end{equation}
with the Macaulay bracket $\langle \bullet \rangle_+ = \max(0, \bullet)$ extracting only tensile contributions.

To ensure the irreversibility of damage, a \emph{history variable}~$\historyvariable \colon [0,\infty) \times \body_0 \times \body_0 \rightarrow \mathbb{R}$ is introduced, tracking the maximum crack driving force experienced by the bond up to time~$t$ as
\begin{equation}
\historyvariable(t,\X,\Xneighbor) = \max_{t'\leq t} \crackdrivingforce(t',\X,\Xneighbor) \eqperiod
\end{equation}
The evolution of the bond phase-field parameter is finally defined as
\begin{equation}
\phasefield(t,\X,\Xneighbor) = \min\!\left(1, \, \frac{\historyvariable(t,\X,\Xneighbor)}{\historyvariable(t,\X,\Xneighbor) + \criticalcrackdrivingforce}\right) \eqcomma
\end{equation}
ensuring that damage can only grow and remains bounded within the interval~$[0,1]$.
This closed-form evolution law provides a smooth damage transition governed by the critical crack driving force~$\criticalcrackdrivingforce$ and guarantees thermodynamic consistency with non-negative crack dissipation.

To connect the bond-wise defined phase-field parameter~$\phasefield$ to a point-wise damage variable similar to the classical peridynamics approach, the bond phase-field parameter is averaged over the family of a material point~$\X$ as
\begin{equation}\label{eq:pointphasefieldavg}
\phasefieldpoint(t,\X) = \int_{\family(\X)} \avgbondkernel(\X,\Xneighbor) \, \phasefield(t,\X,\Xneighbor) \ddVneighbor \eqperiod
\end{equation}
This is identical to the classical definition of a point-wise damage variable in peridynamics, therefore yielding
\begin{equation}\label{eq:pfpdpointdamage}
\pointdamage(t,\X) = \phasefieldpoint(t,\X) \eqperiod
\end{equation}
With this definition, the bond-associated phase-field peridynamics framework seamlessly integrates with existing peridynamic damage modeling approaches and the point-wise damage variable can be used for visualization and post-processing purposes.

\subsection{Determination of the normalization constant}\label{sec:normalizationconstant}

A critical remaining aspect of the bond-associated phase-field peridynamics framework is the determination of the normalization constant~$\normalizationconstant$.
The normalization constant must be chosen such that the total crack dissipation corresponds to Griffith's assumption of energy release rate per unit crack area.

In classical phase-field fracture mechanics, the normalization constant is derived through a rigorous mathematical procedure involving $\Gamma$-convergence arguments.
The approach considers an infinite domain with a one-dimensional damage profile, solves an Euler--Lagrange equation for the optimal damage distribution, and demonstrates convergence to a sharp crack formulation in the limit of vanishing length scale parameter~\cite{Bourdin2000,Bourdin2008}.

In bond-associated phase-field peridynamics, a planar crack in an infinite domain is considered, and the normalization constant is determined by requiring that the total crack dissipation equals the Griffith energy release.
However, the nonlocal nature of the bond-associated peridynamics framework introduces additional geometric considerations, as the crack dissipation depends on the volume of bonds crossing the crack plane weighted by the kernel function.
Therefore, the normalization constant depends on the choice of the kernel function~$\kernel$.

Griffith's assumption states that the energy required to create new crack surface is proportional to the crack area.
For a crack growth of area~$\crackarea$, the required total crack dissipation is
\begin{equation}\label{eq:griffithcrackdissipation}
\totalcrackdissipation = \griffith \, \crackarea \eqperiod
\end{equation}
In the following, the normalization constant is derived for general spherical kernel functions, using \equationref{eq:griffithcrackdissipation} as the defining condition for $\normalizationconstant$.

\subsubsection{Setting and assumptions}

Consider an infinite material domain $\Omega \subset \mathbb{R}^3$ with a planar crack surface~$\cracksurface$ at position $\crackcoordinate = 0$, where the \emph{crack coordinate}~$\crackcoordinate \colon \body_0 \rightarrow \mathbb{R}$ is defined as the signed perpendicular distance from a material point~$\X$ to the crack plane as
\begin{equation}\label{eq:crackcoordinate}
\crackcoordinate(\X) = \X \cdot \cracknormalvector \eqcomma
\end{equation}
with the unit crack normal vector~$\cracknormalvector$.
Points with $\crackcoordinate(\X) > 0$ are located above the crack plane, while points with $\crackcoordinate(\X) < 0$ are located below.
Only points within distance~$\horizon$ of the crack plane have bonds that cross the crack and contribute to the total crack dissipation.

For the derivation, the kernel function is assumed to be spherical, depending only on the bond length~$r := \abs{\Bond}$, i.e., $\kernel(\Bond) = \kernelscalar(r)$ with the scalar kernel function~$\kernelscalar \colon \mathbb{R}^+ \rightarrow \mathbb{R}^+$.
Additionally, the material domain is assumed to be infinite, such that all points involved in the derivation are bulk points with complete spherical neighborhoods.
This implies that the kernel integrals~$\kernelintegral(\X)$ from \equationref{eq:kernelintegraldef} are equal for all material points, i.e., $\kernelintegral(\X) = \kernelintegral(\Xneighbor)$ for all $\X, \Xneighbor \in \Omega$.
Due to the equality, the simplified notation $\kernelintegral := \kernelintegral(\X)$ is used in the following.
Note that the normalization condition from \equationref{eq:kernelintegralnormalization} yields $\kernelintegral = 1$.
However, $\kernelintegral$ is kept explicitly in the following derivation to maintain generality for arbitrary kernel functions.
Under these assumptions, the averaged bond kernel function from \equationref{eq:baqpavgbondkernel} simplifies to
\begin{equation}\label{eq:avgbondkernelsimplified}
\avgbondkernel(\X,\Xneighbor) = \frac{\kernel(\Bond)}{\kernelintegral} \eqperiod
\end{equation}

\subsubsection{Crack dissipation energy density of a material point}

The \emph{point-wise crack dissipation energy density}~$\crackdissipationpoint \colon [0,\infty) \times \body_0 \rightarrow \mathbb{R}$ for a material point~$\X$ is defined as the weighted integral of the bond crack dissipation~$\crackdissipationbond$ from \equationref{eq:bondcrackdissipation} over all neighbors as
\begin{equation}\label{eq:crackdissipationpointdef}
\crackdissipationpoint(t,\X) = \int_{\family(\X)} \avgbondkernel(\X,\Xneighbor) \, \crackdissipationbond(t,\X,\Xneighbor) \ddVneighbor \eqperiod
\end{equation}
The total crack dissipation can be obtained by integrating~$\crackdissipationpoint$ over the material domain.
For a fully propagated sharp crack, all bonds crossing the crack have $\phasefield = 1$, and all bonds not crossing the crack have $\phasefield = 0$.
Consequently, the crack surface density function evaluates to $\cracksurfenergyfunction(\phasefield) = 1$ for crossing bonds and $\cracksurfenergyfunction(\phasefield) = 0$ for non-crossing bonds.

A bond between $\X$ at height $\crackcoordinate(\X)$ and $\Xneighbor$ at height $\crackcoordinate(\Xneighbor)$ crosses the crack if and only if the endpoints are on opposite sides of the crack plane, i.e., $\crackcoordinate(\X) \cdot \crackcoordinate(\Xneighbor) < 0$.
For a point~$\X$ at crack coordinate~$\crackcoordinate = \crackcoordinate(\X) > 0$ (above the crack), all neighbors $\Xneighbor$ with $\crackcoordinate(\Xneighbor) < 0$ (below the crack) form the set of crossing bonds.
This set of neighbors constitutes a \emph{spherical cap}~$\sphericalcap(\X)$, defined as
\begin{equation}\label{eq:sphericalcapdef}
\sphericalcap(\X) = \left\{\Xneighbor \in \family(\X) \; \big| \; \crackcoordinate(\X) \cdot \crackcoordinate(\Xneighbor) < 0 \right\} \eqperiod
\end{equation}
The point-wise crack dissipation then reduces to an integral over the spherical cap as
\begin{equation}\label{eq:crackdissipationpointcap}
\crackdissipationpoint(\X) = \int_{\sphericalcap(\X)} \frac{\kernelscalar(r)}{\kernelintegral} \, \frac{\griffith}{\normalizationconstant \, \horizon} \ddVneighbor = \frac{\griffith}{\normalizationconstant \, \horizon \, \kernelintegral} \int_{\sphericalcap(\X)} \kernelscalar(r) \ddVneighbor \eqperiod
\end{equation}
Since the crack is planar, for any two material points~$\X_1$ and~$\X_2$ with equal crack coordinates $\crackcoordinate(\X_1) = \crackcoordinate(\X_2)$, the spherical caps~$\sphericalcap(\X_1)$ and~$\sphericalcap(\X_2)$ are geometrically congruent.
Combined with the fact that the spherical kernel function~$\kernelscalar$ depends only on the bond length~$r$, the kernel-weighted cap integrals are identical, i.e.,
\begin{equation}
\int_{\sphericalcap(\X_1)} \kernelscalar(r) \ddVneighbor = \int_{\sphericalcap(\X_2)} \kernelscalar(r) \ddVneighbor \eqperiod
\end{equation}
This motivates the definition of the general spherical cap~$\sphericalcapatz(\crackcoordinate) := \sphericalcap(\X)$ for any point $\X\in\Omega$ at a given crack coordinate~$\crackcoordinate = \crackcoordinate(\X)$.
Consequently, the point-wise crack dissipation of all points with equal crack coordinate is identical, i.e., $\crackdissipationpoint(\X_1) = \crackdissipationpoint(\X_2)$.
Therefore, the point-wise crack dissipation can be expressed as a function of the crack coordinate alone, i.e.,
\begin{equation}
\crackdissipationpointatz(\crackcoordinate) = \frac{\griffith}{\normalizationconstant \, \horizon \, \kernelintegral} \int_{\sphericalcapatz(\crackcoordinate)} \kernelscalar(r) \ddVneighbor \eqperiod
\end{equation}

\begin{figure}[ht]
\centering
\includegraphics[scale=\figsizefactor]{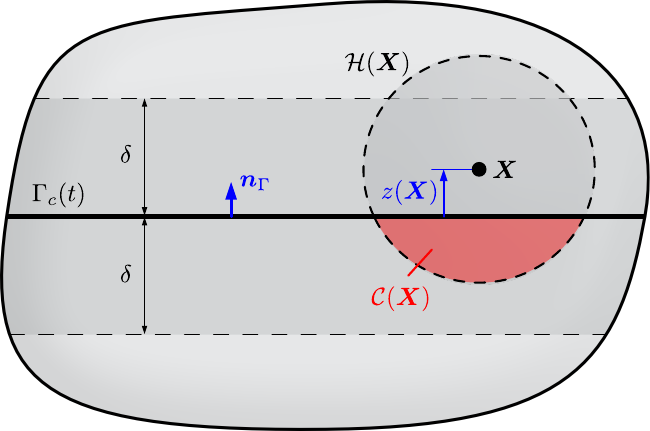}
\caption[Spherical cap geometry for crack dissipation]{Spherical cap geometry for the derivation of the normalization constant. A material point~$\X$ at crack coordinate~$\crackcoordinate = \crackcoordinate(\X) > 0$ has its spherical neighborhood~$\family(\X)$ intersected by the crack plane at $\crackcoordinate = 0$. The spherical cap~$\sphericalcap(\X)$ contains all neighbors below the crack plane whose bonds cross the crack.}
\label{fig:sphericalcapgeometry}
\end{figure}

\subsubsection{Volume of the spherical cap}

The volume of the spherical cap depends mainly on the crack coordinate~$\crackcoordinate(\X)$ of the material point, as illustrated in \autoref{fig:sphericalcapgeometry}.
Since the crack is considered planar and the neighborhood a sphere of radius~$\horizon$ centered at height~$\crackcoordinate$ above the crack plane, the cap volume is given by the standard equation
\begin{equation}\label{eq:capvolumeformula}
\capvolume(\crackcoordinate) = \frac{\pi}{3}(\horizon - \abs{\crackcoordinate})^2(2\horizon + \abs{\crackcoordinate}) \eqperiod
\end{equation}
The absolute value ensures that the symmetry of the cap volume is captured for points below the crack plane with $-\horizon < \crackcoordinate < 0$, enforcing $\capvolume(\crackcoordinate) = \capvolume(-\crackcoordinate)$.
The total volume of the neighborhood sphere is $\familyvolume = \frac{4}{3}\pi\horizon^3$.
The \emph{geometric volume fraction function}~$\volfracfunction \colon [-\horizon,\horizon] \rightarrow [0,\frac{1}{2}]$ is defined as the ratio of the cap volume to the sphere volume, yielding
\begin{equation}
\volfracfunction(\crackcoordinate) = \frac{\capvolume(\crackcoordinate)}{\familyvolume} = \frac{(\horizon - \abs{\crackcoordinate})^2(2\horizon + \abs{\crackcoordinate})}{4\horizon^3} \eqperiod
\end{equation}
This function satisfies $\volfracfunction(0) = \frac{1}{2}$, meaning that a point on the crack plane has exactly half of its neighborhood volume on each side.
Additionally, $\volfracfunction(\pm \horizon) = 0$, indicating that a point at distance~$\horizon$ from the crack has no neighborhood volume on the other side.

Introducing the \emph{normalized crack coordinate}~$\cracknormcoordinate := \abs{\crackcoordinate} / \horizon \in [0,1]$, the volume fraction can be expressed as
\begin{equation}\label{eq:volfracfunctiondef}
\volfracfunction(\cracknormcoordinate) = \frac{\capvolume(\cracknormcoordinate \horizon)}{\familyvolume} = \frac{(\horizon - \cracknormcoordinate\horizon)^2(2\horizon + \cracknormcoordinate\horizon)}{4\horizon^3} = \frac{(1 - \cracknormcoordinate)^2(2 + \cracknormcoordinate)}{4} \qquad \forall \cracknormcoordinate \in [0,1]\eqperiod
\end{equation}
Expanding the numerator leads to the explicit polynomial form
\begin{equation}\label{eq:volfracfunctionexplicit}
\volfracfunction(\cracknormcoordinate) = \frac{1}{2} - \frac{3\cracknormcoordinate}{4} + \frac{\cracknormcoordinate^3}{4} \qquad \forall \cracknormcoordinate \in [0,1] \eqperiod
\end{equation}
Similarly, the normalized form satisfies $\volfracfunction(0) = \frac{1}{2}$ and $\volfracfunction(1) = 0$.
This normalized form will be used in the subsequent derivation of the normalization constant.

\subsubsection{Integration over the crack band}

The total crack dissipation is obtained by integrating the point-wise crack dissipation energy density over the material domain.
Only points within distance~$\horizon$ of the crack contribute, forming a band of thickness~$2\horizon$ centered on the crack plane.
Due to the symmetry of the spherical cap geometry, points at height~$\crackcoordinate$ above the crack and at height~$-\crackcoordinate$ below the crack contribute equally to the total crack dissipation.
The total dissipated energy therefore consists of equal contributions from both sides of the crack.

For the upper block with $0 < \crackcoordinate < \horizon$, the crack dissipation is integrated as follows.
Consider a thin layer of material at height~$\crackcoordinate$ with infinitesimal thickness~$\dd{\crackcoordinate}$ and crack area~$\crackarea$.
The volume of this layer is $\crackarea \dd{\crackcoordinate}$.
Since~$\crackdissipationpoint$ is an energy density (energy per unit volume), the contribution to the total crack dissipation from this layer is
\begin{equation}\label{eq:upperblockcontribution}
\dd{\totalcrackdissipation_{+}} = \crackdissipationpointatz(\crackcoordinate) \, \crackarea \dd{\crackcoordinate} \eqperiod
\end{equation}
The total contribution from the upper block is obtained by integration as
\begin{equation}\label{eq:totalupperblockcontribution}
\totalcrackdissipation_{+} = \crackarea \int_0^{\horizon} \crackdissipationpointatz(\crackcoordinate) \dd{\crackcoordinate} \eqperiod
\end{equation}
Utilizing the normalized crack coordinate~$\cracknormcoordinate$, a \emph{kernel-weighted volume fraction function}~$\kernelcapfrac \colon [0,1] \rightarrow [0,\frac{1}{2}]$ can be defined as
\begin{equation}\label{eq:kernelcapfracdef}
\kernelcapfrac(\cracknormcoordinate) = \frac{1}{\kernelintegral} \int_{\sphericalcapatz(\cracknormcoordinate\horizon)} \kernelscalar(r) \ddVneighbor \eqperiod
\end{equation}
This function gives the fraction of the total kernel weight contained in the spherical cap at height~$\crackcoordinate=\cracknormcoordinate\horizon$.
It satisfies the same boundary conditions as the geometric volume fraction function, i.e., $\kernelcapfrac(0) = \frac{1}{2}$ and $\kernelcapfrac(1) = 0$.

With this definition, the point-wise crack dissipation from \equationref{eq:crackdissipationpointcap} becomes
\begin{equation}\label{eq:crackdissipationpointsimplified}
\crackdissipationpointatz(\crackcoordinate) = \frac{\griffith}{\normalizationconstant \, \horizon} \, \kernelcapfrac(\crackcoordinate/\horizon) \eqperiod
\end{equation}
Inserting this expression into \equationref{eq:totalupperblockcontribution} and changing to the normalized coordinate~$\cracknormcoordinate = \crackcoordinate / \horizon$ with $\dd{\crackcoordinate} = \horizon \, \dd{\cracknormcoordinate}$, the contribution from the upper block yields
\begin{equation}\label{eq:totalupperblock}
\totalcrackdissipation_{+} = \crackarea \, \frac{\griffith}{\normalizationconstant \, \horizon} \int_0^{\horizon} \kernelcapfrac(\crackcoordinate/\horizon) \dd{\crackcoordinate} = \frac{\griffith \, \crackarea}{\normalizationconstant} \int_0^{1} \kernelcapfrac(\cracknormcoordinate) \dd{\cracknormcoordinate} \eqperiod
\end{equation}
Due to the symmetry of the spherical cap geometry, the contribution from the lower block is equal to the upper block contribution.
The total crack dissipation is therefore
\begin{equation}\label{eq:totalcrackdissipationderivation}
\totalcrackdissipation = 2 \, \totalcrackdissipation_{+} = 2 \, \frac{\griffith \, \crackarea}{\normalizationconstant} \int_0^{1} \kernelcapfrac(\cracknormcoordinate) \dd{\cracknormcoordinate} \eqperiod
\end{equation}

\subsubsection{Derivation of the normalization constant}

Applying the Griffith condition from \equationref{eq:griffithcrackdissipation}, the total crack dissipation must equal $\griffith \, \crackarea$.
Substituting this into \equationref{eq:totalcrackdissipationderivation} results in the relation
\begin{equation}\label{eq:griffithconditionapplied}
\griffith \, \crackarea = 2 \, \frac{\griffith \, \crackarea}{\normalizationconstant} \int_0^{1} \kernelcapfrac(\cracknormcoordinate) \dd{\cracknormcoordinate} \eqperiod
\end{equation}
This simplifies to the general expression for the normalization constant as
\begin{equation}\label{eq:normalizationconstantgeneral}
\normalizationconstant = 2 \int_0^{1} \kernelcapfrac(\cracknormcoordinate) \dd{\cracknormcoordinate} \eqperiod
\end{equation}
The factor of~$2$ arises from the geometric symmetry of the crack band, since the total fracture energy includes contributions from both the upper part with $\crackcoordinate \in (0,\horizon)$ and the lower part with $\crackcoordinate \in (-\horizon,0)$.

\subsubsection{Exact evaluation of the normalization constant}

In the following, the general expression for the normalization constant from \equationref{eq:normalizationconstantgeneral} is evaluated exactly.
This requires the exact computation of the kernel-weighted cap fraction~$\kernelcapfrac(\cracknormcoordinate)$ from \equationref{eq:kernelcapfracdef} and subsequently of the integral in \equationref{eq:normalizationconstantgeneral} by reducing the problem to one-dimensional integrals that can be computed analytically or via Gaussian quadrature.

Consider a material point~$\X$ at crack coordinate~$\crackcoordinate = \cracknormcoordinate\horizon > 0$ above the crack plane.
Using spherical coordinates~$(r,\polarangle,\azimuthalangle)$ centered at~$\X$ with the polar axis aligned with the crack normal~$\cracknormalvector$, a neighbor point~$\Xneighbor$ at distance~$r$ and polar angle~$\polarangle$ (measured from the positive crack normal direction) has crack coordinate
\begin{equation}
\crackcoordinate(\Xneighbor) = \cracknormcoordinate \horizon + r \cos(\polarangle) \eqperiod
\end{equation}
The neighbor lies below the crack plane, and thus the bond crosses the crack, if and only if $\crackcoordinate(\Xneighbor) < 0$, which requires
\begin{equation}\label{eq:crossingcondition}
\cos(\polarangle) < -\frac{\cracknormcoordinate \horizon}{r} \eqperiod
\end{equation}
This condition can only be satisfied for bond lengths $r \geq \cracknormcoordinate\horizon$, since $\cos(\polarangle) \geq -1$.
For $r \in [\cracknormcoordinate\horizon, \horizon]$, the polar angle~$\polarangle$ ranges from $\polarangle_0 = \arccos(-\cracknormcoordinate\horizon / r)$ to~$\pi$, while the azimuthal angle~$\azimuthalangle$ ranges over the full interval~$[0,2\pi)$.

The angular part of the volume integral evaluates to
\begin{equation}
\int_{\polarangle_0}^{\pi} \sin(\polarangle) \dd{\polarangle} = \big[-\cos(\polarangle)\big]_{\polarangle_0}^{\pi} = 1 - \frac{\cracknormcoordinate\horizon}{r} \eqperiod
\end{equation}
Combining with the radial and azimuthal integrals, the kernel-weighted cap integral from \equationref{eq:kernelcapfracdef} becomes
\begin{equation}\label{eq:capintegralexact}
\int_{\sphericalcapatz(\cracknormcoordinate\horizon)} \kernelscalar(r) \ddVneighbor = 2\pi \int_{\cracknormcoordinate\horizon}^{\horizon} \kernelscalar(r) \, r^2 \left(1 - \frac{\cracknormcoordinate\horizon}{r}\right) \dd{r} = 2\pi \int_{\cracknormcoordinate\horizon}^{\horizon} \kernelscalar(r) \, r\left(r - \cracknormcoordinate\horizon\right) \dd{r} \eqperiod
\end{equation}
Similarly, the full kernel integral~$\kernelintegral$ from \equationref{eq:kernelintegraldef} for a bulk point with a complete spherical neighborhood evaluates in spherical coordinates to
\begin{equation}\label{eq:kernelintegralspherical}
\kernelintegral = 4\pi \int_0^{\horizon} \kernelscalar(r) \, r^2 \dd{r} \eqperiod
\end{equation}
Introducing the \emph{normalized radial coordinate}~$\rho := r/\horizon \in [0,1]$ with $\dd{r} = \horizon\dd{\rho}$ in both integrals, the factors of~$\pi$ and powers of~$\horizon$ cancel, and the kernel-weighted cap fraction simplifies to the exact expression
\begin{equation}\label{eq:kernelcapfracexact}
\kernelcapfrac(\cracknormcoordinate) = \frac{2\pi\horizon^3 \int_{\cracknormcoordinate}^{1} \kernelscalar(\rho\horizon) \, \rho(\rho - \cracknormcoordinate) \dd{\rho}}{4\pi\horizon^3 \int_0^{1} \kernelscalar(\rho\horizon) \, \rho^2 \dd{\rho}} = \frac{\int_{\cracknormcoordinate}^{1} \kernelscalar(\rho\horizon) \, \rho(\rho - \cracknormcoordinate) \dd{\rho}}{2\int_0^{1} \kernelscalar(\rho\horizon) \, \rho^2 \dd{\rho}} \eqperiod
\end{equation}
This expression is exact for any spherical kernel function.
It can be verified that $\kernelcapfrac(0) = 1/2$ and $\kernelcapfrac(1) = 0$, consistent with the boundary conditions stated in the previous subsubsection.
Moreover, for a constant kernel $\kernelscalar(r) = \mathrm{const.}$, evaluating the integrals yields
\begin{equation}
\kernelcapfrac(\cracknormcoordinate) = \frac{\int_{\cracknormcoordinate}^{1} (\rho^2 - \cracknormcoordinate\rho) \dd{\rho}}{2\int_0^{1} \rho^2 \dd{\rho}} = \frac{\frac{1}{3} - \frac{\cracknormcoordinate}{2} + \frac{\cracknormcoordinate^3}{6}}{\frac{2}{3}} = \frac{1}{2} - \frac{3\cracknormcoordinate}{4} + \frac{\cracknormcoordinate^3}{4} = \volfracfunction(\cracknormcoordinate) \eqcomma
\end{equation}
which recovers the geometric volume fraction from \equationref{eq:volfracfunctionexplicit}, as expected.

Inserting the exact expression for~$\kernelcapfrac$ from \equationref{eq:kernelcapfracexact} into the general formula for the normalization constant from \equationref{eq:normalizationconstantgeneral} yields
\begin{equation}\label{eq:normalizationconstantdoubleintegral}
\normalizationconstant = 2 \int_0^{1} \kernelcapfrac(\cracknormcoordinate) \dd{\cracknormcoordinate} = \frac{\int_0^{1} \int_{\cracknormcoordinate}^{1} \kernelscalar(\rho\horizon) \, \rho(\rho - \cracknormcoordinate) \dd{\rho} \dd{\cracknormcoordinate}}{\int_0^{1} \kernelscalar(\rho\horizon) \, \rho^2 \dd{\rho}} \eqperiod
\end{equation}
The double integral in the numerator can be simplified by exchanging the order of integration.
The integration domain~$\{(\cracknormcoordinate,\rho) \mid 0 \leq \cracknormcoordinate \leq 1, \; \cracknormcoordinate \leq \rho \leq 1\}$ is equivalently described as~$\{(\cracknormcoordinate,\rho) \mid 0 \leq \rho \leq 1, \; 0 \leq \cracknormcoordinate \leq \rho\}$, allowing the order of integration to be switched as
\begin{equation}
\int_0^{1} \int_{\cracknormcoordinate}^{1} \kernelscalar(\rho\horizon) \, \rho(\rho - \cracknormcoordinate) \dd{\rho} \dd{\cracknormcoordinate} = \int_0^{1} \kernelscalar(\rho\horizon) \, \rho \int_0^{\rho} (\rho - \cracknormcoordinate) \dd{\cracknormcoordinate} \dd{\rho} \eqperiod
\end{equation}
The inner integral evaluates analytically to
\begin{equation}
\int_0^{\rho} (\rho - \cracknormcoordinate) \dd{\cracknormcoordinate} = \left[\rho\cracknormcoordinate - \frac{\cracknormcoordinate^2}{2}\right]_0^{\rho} = \rho^2 - \frac{\rho^2}{2} = \frac{\rho^2}{2} \eqperiod
\end{equation}
Therefore, the numerator reduces to a single integral as
\begin{equation}
\int_0^{1} \kernelscalar(\rho\horizon) \, \rho \cdot \frac{\rho^2}{2} \dd{\rho} = \frac{1}{2}\int_0^{1} \kernelscalar(\rho\horizon) \, \rho^3 \dd{\rho} \eqperiod
\end{equation}
Substituting back into \equationref{eq:normalizationconstantdoubleintegral}, the normalization constant is obtained as the exact closed-form expression
\begin{equation}\label{eq:normalizationconstantexact}
\normalizationconstant = \frac{\int_0^{1} \kernelscalar(\rho\horizon) \, \rho^3 \dd{\rho}}{2\int_0^{1} \kernelscalar(\rho\horizon) \, \rho^2 \dd{\rho}} \eqcomma
\end{equation}
in terms of a ratio of two one-dimensional integrals over the unit interval.
Both integrals involve smooth integrands and can be evaluated analytically for polynomial kernels or with high accuracy using Gaussian quadrature.
This result is exact for any spherical kernel function and does not involve any approximations beyond the setting assumptions of an infinite domain with a planar crack.

For the constant kernel $\kernelscalar(r) = 1$, the integrals evaluate to $\int_0^1 \rho^3 \dd{\rho} = 1/4$ and $\int_0^1 \rho^2 \dd{\rho} = 1/3$, yielding the normalization constant
\begin{equation}
\normalizationconstant = \frac{1/4}{2/3} = \frac{3}{8} \eqperiod
\end{equation}
This coincides with the value obtained from integrating the geometric volume fraction function as 
\begin{equation}
\normalizationconstant \, \Big|_{\kernelscalar(r) = 1} = 2\int_0^1 \volfracfunction(\cracknormcoordinate) \dd{\cracknormcoordinate} = 3/8 \eqcomma
\end{equation}
confirming the consistency of the derivation.
The numerical verification of this exact expression for the normalization constant is presented in \autoref{sec:pfpdnormalizationverification}.

\section{Discretization and algorithmic implementation}\label{sec:numerics}

This section presents the spatial discretization of the peridynamic equations introduced in \autoref{sec:peridynamics}, the connection to reproducing kernel methods, and the algorithmic implementation of the PFPD model.

\subsection{Spatial discretization}\label{sec:discretization}

To solve the previously described equations numerically, the continuous body is represented as a discrete set of material points.
A discretized peridynamic body $\bodyd_0 = \left\{\Xd \in \body_0 \; | \; \didx = 1, \, \dots, \, N\right\}$ is defined as a set of $N$ points in the reference domain~$\body_0$.
Each point~$\Xd$ occupies a finite volume~$\Vd > 0$ in the reference configuration.
In a uniform discretization, the points are arranged in a regular grid with a characteristic \emph{point spacing}~$\pointspacing$, with $\Vd = \pointspacing^3$ in three-dimensional space.
The \emph{horizon ratio}~$\horizonratio$ relates the horizon to the point spacing as
\begin{equation}\label{eq:horizonratio}
\horizon = (\horizonratio + \epsilon_\delta) \, \pointspacing \eqcomma
\end{equation}
with the numerical parameter~$\epsilon_\delta$, usually chosen as $\epsilon_\delta = 0.015$, ensuring that the horizon is strictly larger than the point spacing.
The \emph{discrete neighborhood} of a point~$\Xd$ is defined as
\begin{equation}\label{eq:discretefamily}
\familydk := \familyd(\Xd) = \big\{\Xneighbord \in \bodyd_0 \; \big| \; 0 < \abs{\Bondd} \leq \horizon\big\} \eqcomma
\end{equation}
with the discrete bond $\Bondd = \Xneighbord - \Xd$ and neighbor volume~$\Vneighbord > 0$.
In the discrete setting, integrals over the neighborhood are approximated by sums over all neighbors as
\begin{equation}\label{eq:discreteneighborintegral}
\int_{\family(\Xd)} \kernel(\Bond) \, (\bullet) \ddVneighbor \approx \sum_{\Xneighbord \in \familydk} \kernel(\Bondd) \, (\bullet) \, \Vneighbord \eqperiod
\end{equation}
This quadrature rule is applied to all neighborhood integrals, including the shape tensor in \equationref{eq:shapetensor}, the deformation gradient in \equationref{eq:pddefgrad}, the internal force density in \equationref{eq:pdinternalforcedensity}, as well as the strain energy density and force state of the BAQP formulation in \equationref{eq:baqpstrainenergydensity} and \equationref{eq:baqpforcestate}.

\subsection{Reproducing kernel peridynamics notation}\label{sec:rkpd}

The BAQP correspondence formulation can be equivalently expressed in the framework of reproducing kernel (RK) methods~\cite{Bessa2014,Chen2018,Hillman2020}.
This connection arises from the fact that the shape tensor and bond shape vector of the correspondence formulation are special cases of the RK moment matrix and shape function derivatives, respectively~\cite{Bessa2014}.

In the RK framework, the moment matrix is defined as
\begin{equation}\label{eq:rkmomentmatrix}
\rkmomentmatrixn(\Xd) = \sum_{\Xneighbord \in \familydk} \kernel(\Bondd) \, \rkmonomialvecn(\Bondd) \otimes \rkmonomialvecn(\Bondd) \, \Vneighbord \eqcomma
\end{equation}
with the monomial basis vector~$\rkmonomialvecn \in \mathbb{R}^m$ containing $m$ monomial terms up to order $n$.
The RK shape function derivatives are given by
\begin{equation}\label{eq:rkshapefuncderivs}
\dv{\rkshapefunctionn(\Xd)}{\Xd} = \kernel(\Bondd) \, \rkmonomialveczeron \, \rkmomentmatrixn^{-1}(\Xd) \, \rkmonomialvecn(\Bondd) \, \Vneighbord \eqcomma
\end{equation}
where $\rkmonomialveczeron$ denotes the \emph{target derivative matrix} that selects the desired derivative order and direction.
In three-dimensional space, this matrix is defined as
\begin{equation}
\rkmonomialveczeron := \begin{bmatrix}
    \displaystyle\pdv{\rkmonomialvecn(\zerotensor)}{X_1} & \displaystyle\pdv{\rkmonomialvecn(\zerotensor)}{X_2} & \displaystyle\pdv{\rkmonomialvecn(\zerotensor)}{X_3}\\
\end{bmatrix}^{\transpose} \eqcomma
\end{equation}
and contains the partial derivatives of the monomial basis vector evaluated at the zero vector, with each row corresponding to a Cartesian derivative direction.

Choosing the linear monomial basis $\rkmonomialvec_{[1]}(\Bond) = \Bond$ (abbreviated as C1), the target derivative matrix reduces to $\rkmonomialveczeron = \identitytensor$, and the moment matrix reduces exactly to the shape tensor
\begin{equation}
\rkmomentmatrix_{[1]}(\Xd) = \shapetensor(\Xd) \eqcomma
\end{equation}
and the shape function derivatives become
\begin{equation}
\dv{\rkshapefunctionn(\Xd)}{\Xd} = \bondshapevector(\Xd, \Xneighbord) \, \Vneighbord \eqperiod
\end{equation}
This reveals that the correspondence formulation is inherently a first-order accurate reproducing kernel approximation.
Higher-order accuracy can be achieved by enriching the monomial basis, for instance with the RK1 basis $\rkmonomialvec_{[1]}(\Bond) = [1 , \scalarBond_1 , \scalarBond_2 , \scalarBond_3]\transpose$ that includes a constant term to enforce partition of unity.
With the RK1 basis, the target derivative matrix in three-dimensional space becomes
\begin{equation}
\rkmonomialveczeron = \begin{bmatrix}
	0 & 1 & 0 & 0\\
	0 & 0 & 1 & 0\\
	0 & 0 & 0 & 1\\
\end{bmatrix} \eqperiod
\end{equation}
Using an enriched basis improves the approximation quality, especially near boundaries where the neighborhood is truncated~\cite{Hillman2020,Bode2020}.

In the RK notation, the deformation gradient from \equationref{eq:pddefgradbondshapevector} is expressed as
\begin{equation}
\pddefgrad(t,\Xd) = \identitytensor + \sum_{\Xneighbord \in \familydk} \DispBondd \otimes \dv{\rkshapefunctionn(\Xd)}{\Xd} = \sum_{\Xneighbord \in \familydk} \bond \otimes \dv{\rkshapefunctionn(\Xd)}{\Xd}\eqcomma
\end{equation}
where $\DispBondd = \Displacement(t,\Xneighbord) - \Displacement(t,\Xd)$ is the bond displacement.
The algorithms below use this notation, since it naturally accommodates the kinematic degradation required for the phase-field approach and generalizes to higher-order bases.

\subsection{Discrete PFPD formulation and algorithms}

To write the equations in a discretized form, the \emph{discrete bond phase-field parameter} is introduced as
\begin{equation}
\phasefieldd := \phasefield(t,\Xd,\Xneighbord) \eqcomma
\end{equation}
with the discrete material points $\Xd$ and $\Xneighbord\in\familydk$ in a discretized body $\bodyd_0$.
Analogously, the \emph{discrete history variable} and the \emph{discrete crack driving force} are defined as bond-wise quantities for each pair of discrete material points as
\begin{equation}
\historyvariabled := \historyvariable(t,\Xd,\Xneighbord) \eqcomma \qquad \crackdrivingforcedk := \crackdrivingforce(t,\Xd,\Xneighbord) \eqperiod
\end{equation}
The \emph{discrete point damage} is the point-wise damage variable obtained from averaging the bond phase-field parameters over the family of~$\Xd$, as defined in \equationref{eq:pfpdpointdamage}, yielding
\begin{equation}
\pointdamaged := \pointdamage(t,\Xd) \eqperiod
\end{equation}

Analogously to the degradation of the shape tensor and bond shape vector in \equationref{eq:pfpdshapetensor} and \equationref{eq:pfpdbondshapevector}, the PFPD framework can be expressed using a reproducing kernel peridynamics (RKPD) notation, cf.~\cite{Bessa2014,Chen2018}.
The shape tensor and bond shape vector are replaced by the moment matrix and shape function derivatives in the reproducing kernel setting.
The moment matrix is degraded by the bond phase-field parameter as
\begin{equation}
\rkmomentmatrixn(t,\Xd) = \sum_{\Xneighbord \in \familydk} \kernel(\Bondd) \, \kindegradation(\phasefieldd) \, \rkmonomialvecn(\Bondd) \otimes \rkmonomialvecn(\Bondd) \, \Vneighbord \eqcomma
\end{equation}
where the kinematic degradation function~$\kindegradation$ uses the discrete bond phase-field parameter~$\phasefieldd$ and the critical phase-field value~$\phasefieldcrit$, as defined in \equationref{eq:pfpdkindegradation}.
The shape function derivatives are then degraded similarly as
\begin{align}
\dv{\rkshapefunctionn(t,\Xd)}{\Xd} = \kernel(\Bondd) \, \kindegradation(\phasefieldd) \, \rkmonomialveczeron \, \rkmomentmatrixn^{-1}(t,\Xd) \, \rkmonomialvecn(\Bondd) \, \Vneighbord \eqperiod
\end{align}
This kinematic degradation ensures that the inversion of the moment matrix remains stable even for finite deformations and largely damaged bonds, while maintaining the capability for full removal of the energetic contribution of fully damaged bonds with $\phasefieldd = 1$.
Due to the delayed application of the kinematic degradation with $\phasefieldcrit$, the provided scheme is numerically much more stable than a direct degradation of the moment matrix with the same degradation function as the energetic contribution.

To highlight the differences between the standard BAQP model with critical stretch damage and the novel phase-field peridynamics (PFPD) model, both approaches are presented as algorithms for the evaluation of the internal force density~$\bint$ within a single time step.
The time integration is handled by a standard explicit time integration scheme such as Velocity Verlet~\cite{Silling2005}, and is identical for both models.
The algorithmic differences are entirely contained in the force evaluation procedure.

In the standard \textbf{BAQP model}, fracture is modeled by deleting bonds whose bond strain exceeds a critical value~$\criticalbondstrain$.
A deleted bond carries no force and is excluded from the kinematic approximation.
The evaluation of~$\bint$ proceeds in four steps, as described in \autoref{alg:baqp}.
In Step~0, the bond strain~$\bondstrain$ is evaluated for each bond and compared against the critical value.
Bonds exceeding the threshold are deactivated by setting the failure flag~$\bondfailured$ to zero, and the point damage~$\pointdamaged$ is updated as the volume-weighted fraction of failed bonds.
In Step~1, the moment matrix~$\rkmomentmatrixn$ and the shape function derivatives are reassembled for all points where the damage state changed, using only the remaining active bonds with $\bondfailured = 1$.
Step~2 assembles the nonlocal deformation gradient~$\pddefgrad$ for all points.
In Step~3, the bond-associated deformation gradient~$\pdbonddefgrad$ is computed for each active bond as the average of the nonlocal deformation gradients at both endpoints, corrected to reproduce the actual bond deformation exactly.
The first Piola-Kirchhoff stress tensor~$\pdbondfirstpiola$ is then evaluated from the constitutive law.
The force state~$\forcestate$ is assembled from two contributions: the direct bond stress projection along the bond direction and the non-uniform stress correction~$\baqpnonunistresstensor$ distributed via the shape function derivatives.
The internal force density~$\bint$ is accumulated by summing the force state contributions over all active bonds.

In the \textbf{PFPD model}, bonds are not deleted when the critical crack driving force is exceeded.
Instead, all bonds remain part of the kinematic approximation, and the stress response is continuously degraded via the degradation function~$\degradation(\phasefield)$.
The evaluation of~$\bint$ follows the same four-step structure, as described in \autoref{alg:pfpd}, but differs in several key aspects.
In Step~0, the point damage~$\pointdamaged$ is computed from the bond phase-field parameters of the previous time step using the kernel-weighted averaging from \equationref{eq:pfpdpointdamage}, replacing the volume-weighted failure fraction of the BAQP model.
In Step~1, the moment matrix is updated for all points where any bond phase-field parameter exceeds the critical value~$\phasefieldcrit$.
Notably, the binary failure flag~$\bondfailured$ of the BAQP model is replaced by the continuous kinematic degradation function~$\kindegradation(\phasefieldd)$, which ensures a gradual reduction of the kinematic contribution for highly damaged bonds while preserving the stability of the moment matrix.
Step~2 is identical to the BAQP model.
The main difference lies in Step~3: the bond-associated deformation gradient is computed for \emph{all} bonds, regardless of their damage state.
The undamaged first Piola-Kirchhoff stress tensor~$\undamagedfirstpiola$ is evaluated first from the constitutive law.
Subsequently, the crack driving force~$\crackdrivingforcedk$ is determined from the tensile strain energy density or the maximum principal stress, as introduced in \equationref{eq:crackdrivingforcestrainenergy} and \equationref{eq:crackdrivingforcestress}, respectively.
The history variable~$\historyvariabled$ is updated by taking the maximum of its previous value and the current crack driving force, and the bond phase-field parameter~$\phasefieldd$ is evolved accordingly.
The degraded stress tensor $\pdbondfirstpiola = \degradation(\phasefieldd) \, \undamagedfirstpiola$ is then used to assemble the non-uniform stress correction and the internal force density, following the same procedure as in the BAQP model.

The key differences between both algorithms are summarized as follows.
In the BAQP model (\autoref{alg:baqp}), bonds are abruptly deleted when the critical strain is exceeded, and the moment matrix is reassembled using only the remaining active bonds.
This binary bond deletion progressively degrades the kinematic approximation and can lead to numerical instabilities, especially in regions with many deleted bonds.
Moreover, the stress response of a bond transitions instantaneously from undamaged to fully inactive, which does not allow for a smooth dissipation of energy during crack propagation.
In contrast, the PFPD model (\autoref{alg:pfpd}) retains all bonds and continuously degrades the stress contribution of damaged bonds via the degradation function~$\degradation(\phasefield)$.
The kinematic approximation is only gradually affected through~$\kindegradation(\phasefield)$, preserving the stability of the moment matrix even under large-scale fracture.
The phase-field evolution via the history variable~$\historyvariable$ ensures irreversibility and thermodynamic consistency of the damage process.
Another notable difference concerns the input and output quantities of the algorithms: while the BAQP model operates on binary bond failure flags~$\bondfailured$, the PFPD model requires the continuous history variable~$\historyvariabled$ and the bond phase-field parameter~$\phasefieldd$ to be stored and updated for each bond at every time step.

\begin{algorithm}[!ht]
\DontPrintSemicolon
\caption[BAQP with critical stretch, evaluation of the internal force density]{BAQP with critical stretch, evaluation of $\bint$}\label{alg:baqp}
\KwIn{Positions $\Xd$, $\xd$; bond failure $\bondfailured$ from previous time step}
\KwOut{Force density $\bint(\Xd)$; updated bond failure $\bondfailured$}
\BlankLine
\jlcomment{Step 0: Bond failure}
\ForEach{point $\Xd \in \bodyd$}{
	\ForEach{bond $\Bondd$ with $\Xneighbord \in \familydk$}{
		$\displaystyle\bondstrain(\Xd,\Xneighbord) \leftarrow \frac{\abs{\bondd} - \abs{\Bondd}}{\abs{\Bondd}}$\;
		\If{$\bondstrain(\Xd,\Xneighbord) > \criticalbondstrain$ and failure is permitted}{
			Deactivate bond: $\bondfailured \leftarrow 0$\;
		}
	}
}
\BlankLine
\jlcomment{Step 1: Moment matrix and shape function derivatives}
\ForEach{point $\Xd \in \bodyd$ where damage changed}{
	$\displaystyle\rkmomentmatrixn(\Xd) \leftarrow \sum_{\Xneighbord \in \familydk} \kernel(\Bondd) \, \bondfailured \, \rkmonomialvecn(\Bondd) \otimes \rkmonomialvecn(\Bondd) \, \Vneighbord$\;
	\ForEach{bond $\Bondd$ with $\Xneighbord \in \familydk$}{
		$\displaystyle\dv{\rkshapefunctionn(\Xd)}{\Xd} \leftarrow \kernel(\Bondd) \, \bondfailured \, \rkmonomialveczeron \, \rkmomentmatrixn^{-1}(\Xd) \, \rkmonomialvecn(\Bondd) \, \Vneighbord \eqperiod$\;
	}
}
\BlankLine
\jlcomment{Step 2: Nonlocal deformation gradient}
\ForEach{point $\Xd \in \bodyd$}{
	$\displaystyle\pddefgrad(\Xd) \leftarrow \identitytensor + \sum_{\Xneighbord \in \familydk} \DispBondd \otimes \dv{\rkshapefunctionn(\Xd)}{\Xd}$\;
}
\BlankLine
\jlcomment{Step 3: Bond stress and force density}
\ForEach{point $\Xd \in \bodyd$}{
	Initialize $\baqpnonunistresstensor(\Xd) \leftarrow \zerotensor$\;
	Initialize $\bint(\Xd) \leftarrow \zerotensor$\;
	\ForEach{active bond $\Bondd$ with $\Xneighbord \in \familydk$ and $\bondfailured = 1$}{
		$\pdbonddefgrad^{\mathrm{av}}(\Xd,\Xneighbord) \leftarrow \frac{1}{2} \left(\pddefgrad(\Xd) + \pddefgrad(\Xneighbord)\right)$\;
		$\displaystyle\pdbonddefgrad(\Xd,\Xneighbord)  \leftarrow \pdbonddefgrad^{\mathrm{av}}(\Xd,\Xneighbord) + \left(\bondd- \pdbonddefgrad^{\mathrm{av}}(\Xd,\Xneighbord) \, \Bondd\right) \otimes \frac{\Bondd}{\abs{\Bondd}^2}$\;
		$\pdbondfirstpiola(\Xd,\Xneighbord) \leftarrow \firstpiola(\pdbonddefgrad(\Xd,\Xneighbord))$\;
		$\displaystyle\baqpnonunistresstensor(\Xd) \leftarrow \baqpnonunistresstensor(\Xd) + \avgbondkernel(\Xd,\Xneighbord) \, \pdbondfirstpiola(\Xd,\Xneighbord) \left(\identitytensor - \frac{\Bondd \otimes \Bondd}{\abs{\Bondd}^2}\right) \Vneighbord$ \;
	}
	\ForEach{active bond $\Bondd$ with $\Xneighbord \in \familydk$ and $\bondfailured = 1$}{
		$\displaystyle\forcestate(\Xd,\Xneighbord) \leftarrow \avgbondkernel(\Xd,\Xneighbord) \, \pdbondfirstpiola(\Xd,\Xneighbord) \frac{\Bondd}{\abs{\Bondd}} + \baqpnonunistresstensor(\Xd) \dv{\rkshapefunctionn(\Xd)}{\Xd} \frac{1}{\Vneighbord}$\;
		$\bint(\Xd) \leftarrow \bint(\Xd) + \forcestate(\Xd,\Xneighbord) \, \Vneighbord$\;
		$\bint(\Xneighbord) \leftarrow \bint(\Xneighbord) - \forcestate(\Xd,\Xneighbord) \, \Vd$\;
	}
}
\end{algorithm}

\begin{algorithm}[!ht]
\DontPrintSemicolon
\caption[PFPD with phase-field damage, evaluation of the internal force density]{PFPD with phase-field damage, evaluation of $\bint$}\label{alg:pfpd}
\KwIn{Positions $\Xd$, $\xd$; history $\historyvariabled$; phase-field $\phasefieldd$ from previous time step}
\KwOut{Force density $\bint(\Xd)$; updated history~$\historyvariabled$ and phase-field~$\phasefieldd$}
\BlankLine
\jlcomment{Step 1: Moment matrix and shape function derivatives}
\ForEach{point $\Xd \in \bodyd$ where $\phasefieldd > \phasefieldcrit$}{
	$\displaystyle\rkmomentmatrixn(\Xd) \leftarrow \sum_{\Xneighbord \in \familydk} \kernel(\Bondd) \, \kindegradation(\phasefieldd) \, \rkmonomialvecn(\Bondd) \otimes \rkmonomialvecn(\Bondd) \, \Vneighbord$\;
	\ForEach{bond $\Bondd$ with $\Xneighbord \in \familydk$}{
		$\displaystyle\dv{\rkshapefunctionn(\Xd)}{\Xd} \leftarrow \kernel(\Bondd) \, \kindegradation(\phasefieldd) \, \rkmonomialveczeron \, \rkmomentmatrixn^{-1}(\Xd) \, \rkmonomialvecn(\Bondd) \, \Vneighbord \eqperiod$\;
	}
}
\BlankLine
\jlcomment{Step 2: Nonlocal deformation gradient}
\ForEach{point $\Xd \in \bodyd$}{
	$\displaystyle\pddefgrad(\Xd) \leftarrow \identitytensor + \sum_{\Xneighbord \in \familydk} \DispBondd \otimes \dv{\rkshapefunctionn(\Xd)}{\Xd}$\;
}
\BlankLine
\jlcomment{Step 3: Bond stress, damage update, and force density}
\ForEach{point $\Xd \in \bodyd$}{
	Initialize $\baqpnonunistresstensor(\Xd) \leftarrow \zerotensor$\;
    Initialize $\bint(\Xd) \leftarrow \zerotensor$\;
	\ForEach{bond $\Bondd$ with $\Xneighbord \in \familydk$}{
		$\pdbonddefgrad^{\mathrm{av}}(\Xd,\Xneighbord) \leftarrow \frac{1}{2} \left(\pddefgrad(\Xd) + \pddefgrad(\Xneighbord)\right)$\;
		$\displaystyle\pdbonddefgrad(\Xd,\Xneighbord)  \leftarrow \pdbonddefgrad^{\mathrm{av}}(\Xd,\Xneighbord) + \left(\bondd- \pdbonddefgrad^{\mathrm{av}}(\Xd,\Xneighbord) \, \Bondd\right) \otimes \frac{\Bondd}{\abs{\Bondd}^2}$\;
		$\undamagedfirstpiola(\Xd,\Xneighbord) \leftarrow \firstpiola(\pdbonddefgrad(\Xd,\Xneighbord))$\;
		$\crackdrivingforcedk \leftarrow \tensilestrainenergydensity(\pdbonddefgrad(\Xd,\Xneighbord))$ or $\displaystyle\frac{\langle\sigma_1(\pdbonddefgrad(\Xd,\Xneighbord))\rangle_+^2}{2 \, \youngsmodulus}$\;
		$\historyvariabled \leftarrow \max\!\left(\historyvariabled, \, \crackdrivingforcedk\right)$\;
		$\phasefieldd \leftarrow \min\!\left(1, \; \historyvariabled \, / \, (\historyvariabled + \criticalcrackdrivingforce)\right)$\;
		$\pdbondfirstpiola(\Xd,\Xneighbord) \leftarrow \degradation(\phasefieldd) \, \undamagedfirstpiola(\Xd,\Xneighbord)$\;
		$\displaystyle\baqpnonunistresstensor(\Xd) \leftarrow \baqpnonunistresstensor(\Xd) + \avgbondkernel(\Xd,\Xneighbord) \, \pdbondfirstpiola(\Xd,\Xneighbord) \left(\identitytensor - \frac{\Bondd \otimes \Bondd}{\abs{\Bondd}^2}\right) \Vneighbord$ \;
	}
	\ForEach{bond $\Bondd$ with $\Xneighbord \in \familydk$}{
		$\displaystyle\forcestate(\Xd,\Xneighbord) \leftarrow \avgbondkernel(\Xd,\Xneighbord) \, \pdbondfirstpiola(\Xd,\Xneighbord) \frac{\Bondd}{\abs{\Bondd}} + \baqpnonunistresstensor(\Xd) \dv{\rkshapefunctionn(\Xd)}{\Xd} \frac{1}{\Vneighbord}$\;
		$\bint(\Xd) \leftarrow \bint(\Xd) + \forcestate(\Xd,\Xneighbord) \, \Vneighbord$\;
		$\bint(\Xneighbord) \leftarrow \bint(\Xneighbord) - \forcestate(\Xd,\Xneighbord) \, \Vd$\;
	}
}
\end{algorithm}

\section{Numerical Examples}\label{sec:examples}

The capabilities of the PFPD model are demonstrated through a series of numerical examples.
First, mode~I and mode~II crack propagation in a notched plate are investigated to verify the correct crack path prediction.
Then, the boundary tension test (BTT) is used to verify the normalization constant derived in \autoref{sec:normalizationconstant} and to demonstrate the ability of the PFPD model to capture crack branching.
Finally, the Kalthoff-Winkler experiment is simulated to assess the performance of the model under impact-induced fracture conditions.
The simulations are performed in three dimensions using Velocity Verlet time integration~\cite{Silling2005}. 
If not stated otherwise, a critical phase-field value of $\phasefieldcrit = 0.95$, and the Saint Venant-Kirchhoff constitutive model with linear elastic material parameters are employed.
The open-source Julia package \texttt{Peridynamics.jl}~\cite{Partmann2024JuliaCon,Partmann2025_Peridynamics} is used for all simulations.
The PFPD formulation presented in this work is currently implemented in a private development branch of the package and will be released in a future version.

\subsection{Mode~I and mode~II crack propagation}\label{sec:mode_i_ii}

As a first verification of the PFPD model, mode~I and mode~II crack propagation in a notched plate are investigated.
A rectangular plate with dimensions $L_{xy} \times L_{xy} \times \frac{1}{10} L_{xy}$ with $L_{xy} = \qty{100}{mm}$ and a pre-defined central notch of length $\frac{1}{2} L_{xy}$ is considered.
The plate is uniformly discretized with point spacing $\pointspacing = L_{xy} / 120$, yielding a $120 \times 120 \times 12$ point cloud with $N = \num{172800}$ material points.
The pre-notch is introduced by breaking all bonds crossing the notch surface before the simulation.
One layer of boundary points at the top and bottom surfaces is defined as a no-failure zone.
The material parameters are density $\densityref = \qty{2500}{kg/m^3}$, Young's modulus $\youngsmodulus = \qty{32000}{MPa}$, Poisson's ratio $\poissonsratio = 0.25$, and critical energy release rate $\griffith = \qty{100}{J/m^2}$.

\begin{figure}[ht]
\centering
\begin{minipage}{0.49\textwidth}
\centering
\includegraphics[width=\figsize{0.49\textwidthdissertation}]{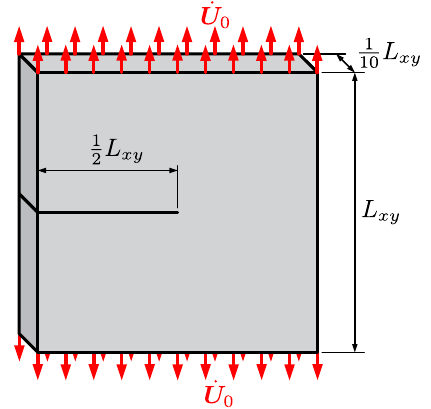}
\end{minipage}%
\hfill
\begin{minipage}{0.49\textwidth}
\centering
\includegraphics[width=\figsize{0.49\textwidthdissertation}]{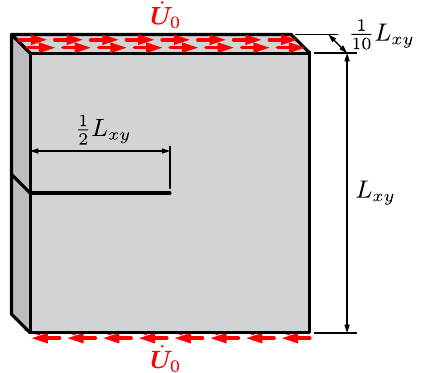}
\end{minipage}
\caption{Setup for the mode~I (left) and mode~II (right) crack propagation examples}
\label{fig:mode_setup}
\end{figure}

\begin{figure}[ht]
\centering
\begin{minipage}{0.49\textwidth}
\centering
\hspace*{3pt} PFPD mode~I\\
\includegraphics[width=\textwidth]{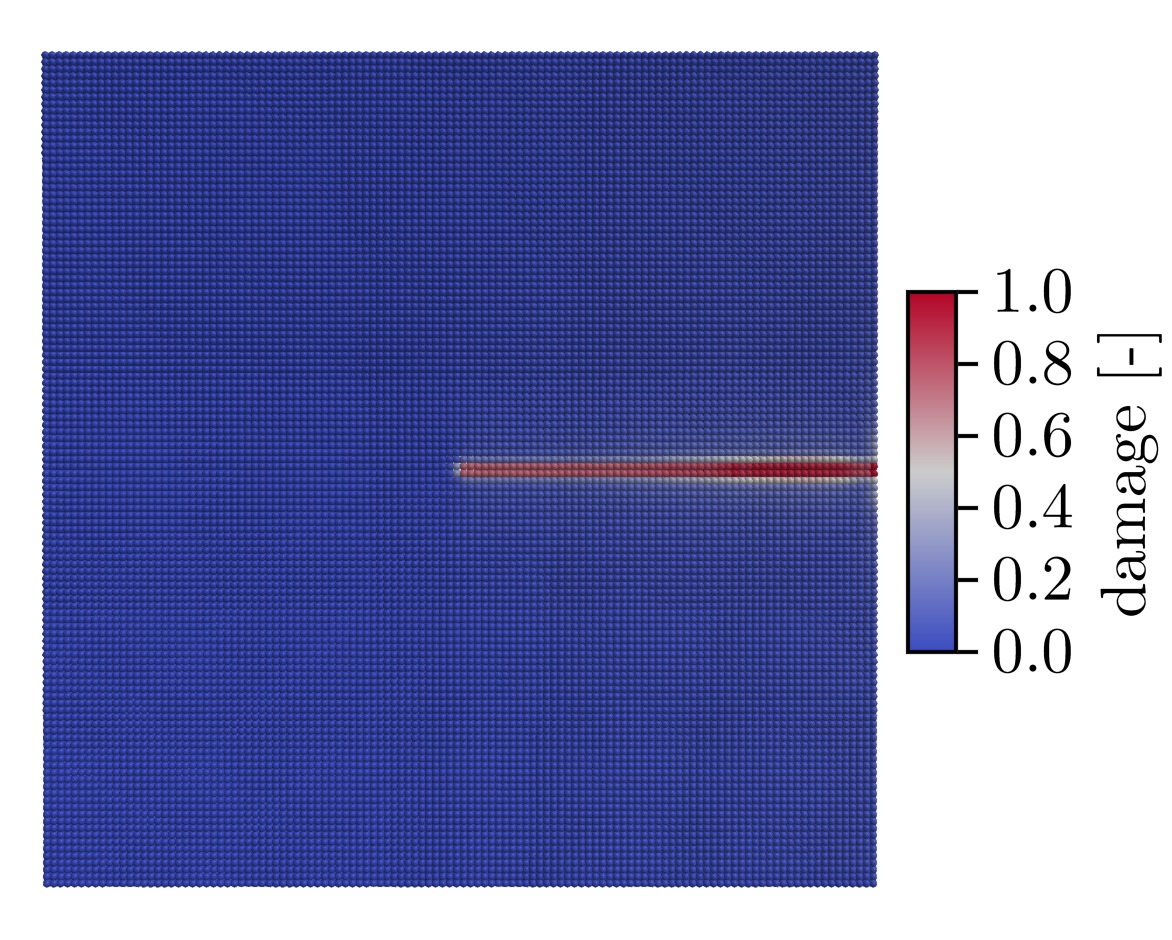}
\end{minipage}%
\hfill
\begin{minipage}{0.49\textwidth}
\centering
\hspace*{3pt} PFPD mode~II\\
\includegraphics[width=\textwidth]{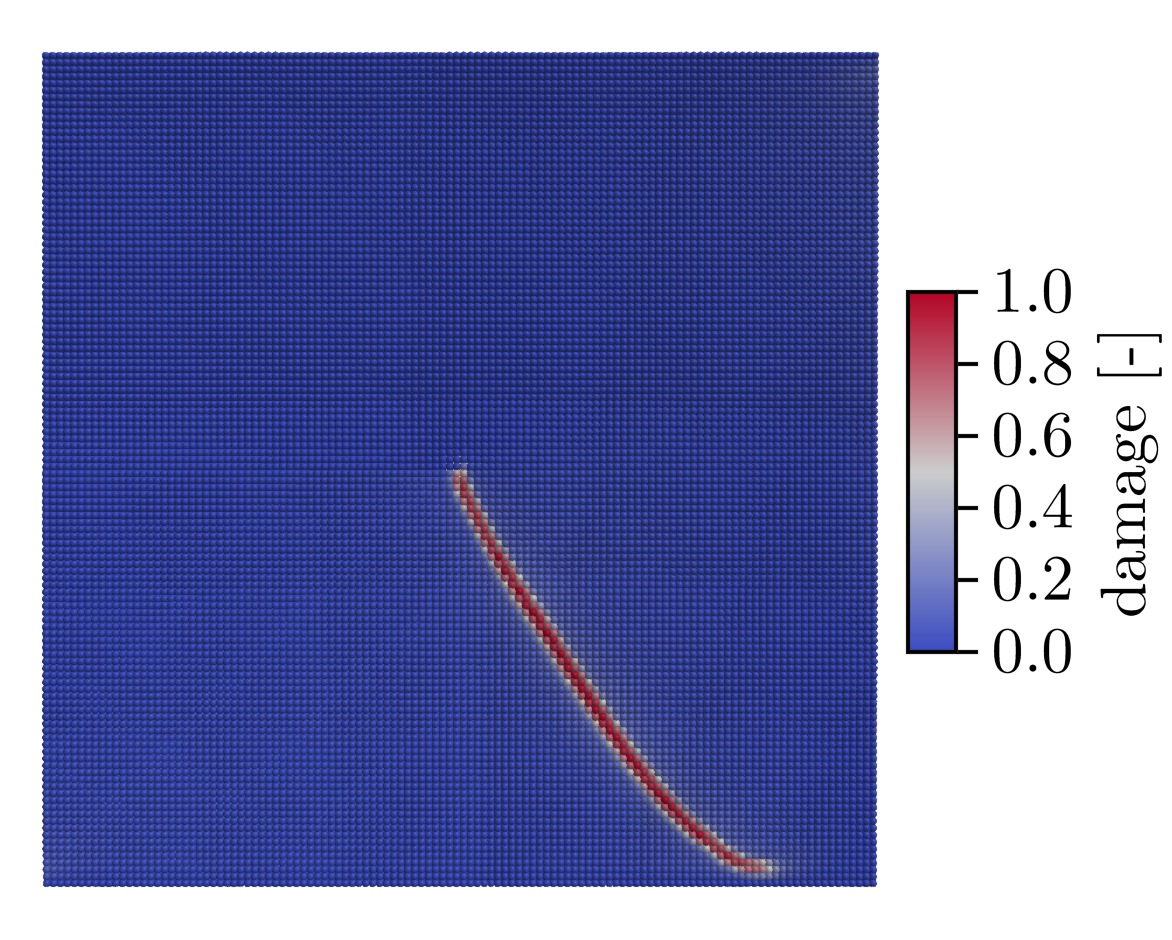}
\end{minipage}
\caption{Mode~I and mode~II crack propagation in 3D at time $t = \qty{171}{\micro\second}$ for the PFPD model}
\label{fig:mode_i_ii_results_pfpd}
\end{figure}

The loading is imposed via a Dirichlet boundary condition~$\Velocity_{\mathrm{D}}$ on the upper and lower boundaries as
\begin{equation}\label{eq:mode_i_ii_dirichlet}
\Velocity(t,\X) = \Velocity_{\mathrm{D}}(t,\X) \qquad \forall \X \in \body_0 \cap \dirichletboundary \eqcomma
\end{equation}
where $\dirichletboundary$ denotes the set of points on the upper and lower boundaries, and the prescribed boundary velocity is constant with $\Velocity_{\mathrm{D}}(t,\X) = \Velocity_0$.
For mode~I, a Dirichlet boundary condition with $\Velocity_0 = \pm \dot{U}_0 \, \vb{e}_2$ and $\dot{U}_0 = \qty{0.2}{m/s}$ is applied at the upper and lower boundaries, pulling the plate apart symmetrically, as shown in \autoref{fig:mode_setup} (left).
For mode~II, the loading is changed to shear: the upper and lower halves are displaced in opposite horizontal directions with $\Velocity_0 = \pm \dot{U}_0 \, \vb{e}_1$ and $\dot{U}_0 = \qty{0.1}{m/s}$, as depicted in \autoref{fig:mode_setup} (right).
All other geometric and material parameters remain identical.

In \autoref{fig:mode_i_ii_results_pfpd}, the crack patterns obtained with the PFPD model are shown.
For mode~I, the crack propagates as a straight horizontal line from the initial notch tip, which is the expected behavior for symmetric tensile loading.
For mode~II, the crack deviates from the initial notch direction and follows a curved path, consistent with the expected behavior for shear-dominated fracture.
Both results are in agreement with results from standard bond-based and other correspondence peridynamics formulations, confirming that the PFPD model correctly reproduces the expected crack paths for both loading modes.
Notably, the standard BAQP model with critical stretch damage becomes unstable in this mode~II case due to shape tensor degradation: the asymmetric bond deletion under shear loading leads to an ill-conditioned shape tensor and eventually to numerical failure.
The PFPD model, in contrast, remains stable throughout the simulation, since bonds are not deleted but continuously degraded via the degradation function.

\subsection{Verification of the normalization constant}\label{sec:pfpdnormalizationverification}

\begin{figure}[ht]
\centering
\includegraphics[width=\figsize{0.7\textwidthdissertation}]{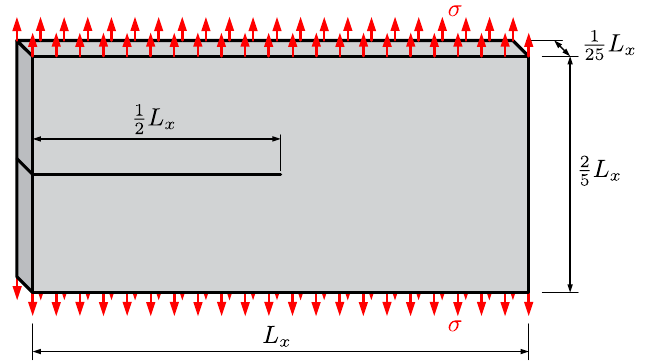}
\caption{Setup of the boundary tension test (BTT)}
\label{fig:btt_setup}
\end{figure}

To verify the correctness of the normalization constant~$\normalizationconstant$ derived in \autoref{sec:normalizationconstant}, the boundary tension test (BTT) is employed.
The BTT, originally introduced by Bourdin et al.~\cite{Bourdin2008} for phase-field fracture models, is a standard benchmark for dynamic crack branching.
A rectangular plate with dimensions $L_x \times \frac{2}{5} L_x \times \frac{1}{25} L_x$ and $L_x = \qty{100}{mm}$ is considered.
The plate is uniformly discretized with $N_x \times N_y \times N_z$ points and point spacing $\pointspacing = \frac{2}{5} L_x / N_y$.
A pre-defined crack extends from the left side to the center of the plate.
The material parameters are density $\densityref = \qty{2450}{kg/m^3}$, Young's modulus $\youngsmodulus = \qty{32000}{MPa}$, Poisson's ratio $\poissonsratio = 0.25$, and critical energy release rate $\griffith = \qty{3}{J/m^2}$.
The setup is shown in \autoref{fig:btt_setup}.

The loading is applied through Neumann boundary conditions $\bextN(t, \X)$ at the upper and lower boundaries with
\begin{equation}
\bext(t, \X) = \bextN(t, \X) \qquad \forall \X \in \body_0 \cap \neumannboundary \eqcomma
\end{equation}
where $\neumannboundary$ denotes one layer of points on the upper and lower boundaries.
A constant dynamic stress of $\sigma = \qty{1}{MPa}$ is applied as body force with 
\begin{equation}
\bextN(t, \X) = \frac{\pm \sigma}{\pointspacing} \, \vb{e}_2 \eqcomma
\end{equation}
at the upper and lower boundaries, respectively.

If the normalization constant is correct, the critical crack driving force~$\criticalcrackdrivingforce$ from \equationref{eq:criticalcrackdrivingforcestrainenergy} yields consistent fracture behavior regardless of the choice of kernel function and horizon ratio.
To this end, the BTT is simulated with different horizon ratios $\horizonratio \in \{3,4,5,6\}$ for two different kernel functions, each with a discretization of $N_y = 100$.

\subsubsection{Results with the cubic B-spline kernel}

\begin{figure}[!ht]
\centering
\includegraphics[width=0.99\textwidth]{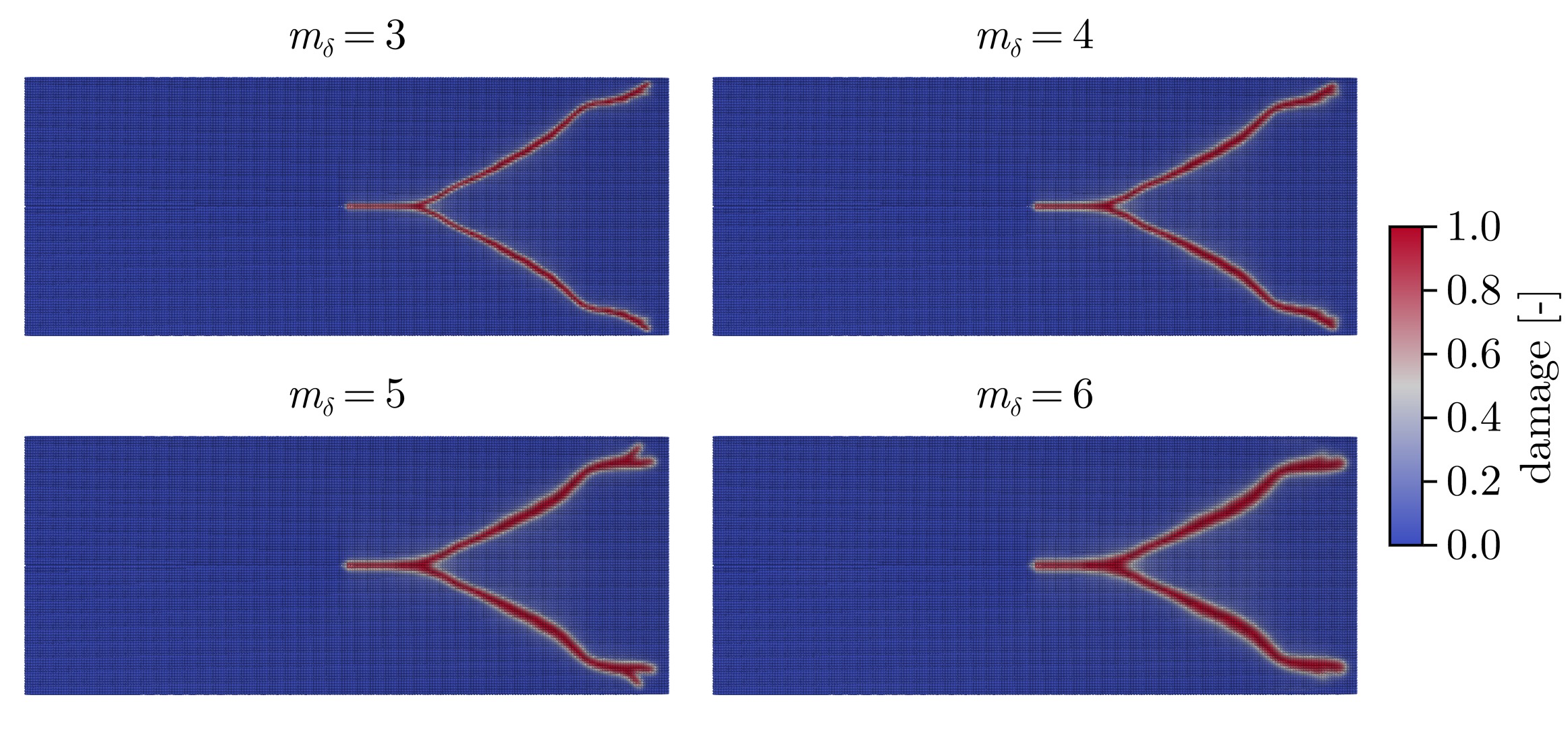}
\caption{Crack branching patterns for different horizon ratios in the BTT for the PFPD model with the cubic B-spline kernel from \equationref{eq:baqpcubicsplinekernel}}
\label{fig:bttbaqppf_k3_c0verification}
\end{figure}

\autoref{fig:bttbaqppf_k3_c0verification} shows the crack branching patterns obtained with the cubic B-spline kernel from \equationref{eq:baqpcubicsplinekernel}.
All four simulations produce very similar crack paths, with the same branching point and comparable branching angles.
As expected, higher horizon ratios lead to a thicker damage zone, which is a direct consequence of the increased nonlocal interaction radius.
The same results are obtained for different discretization values $N_y$, confirming the consistency of the results and the correctness of the normalization constant for this kernel function.

\begin{figure}[!ht]
\centering
\includegraphics[width=0.99\textwidth]{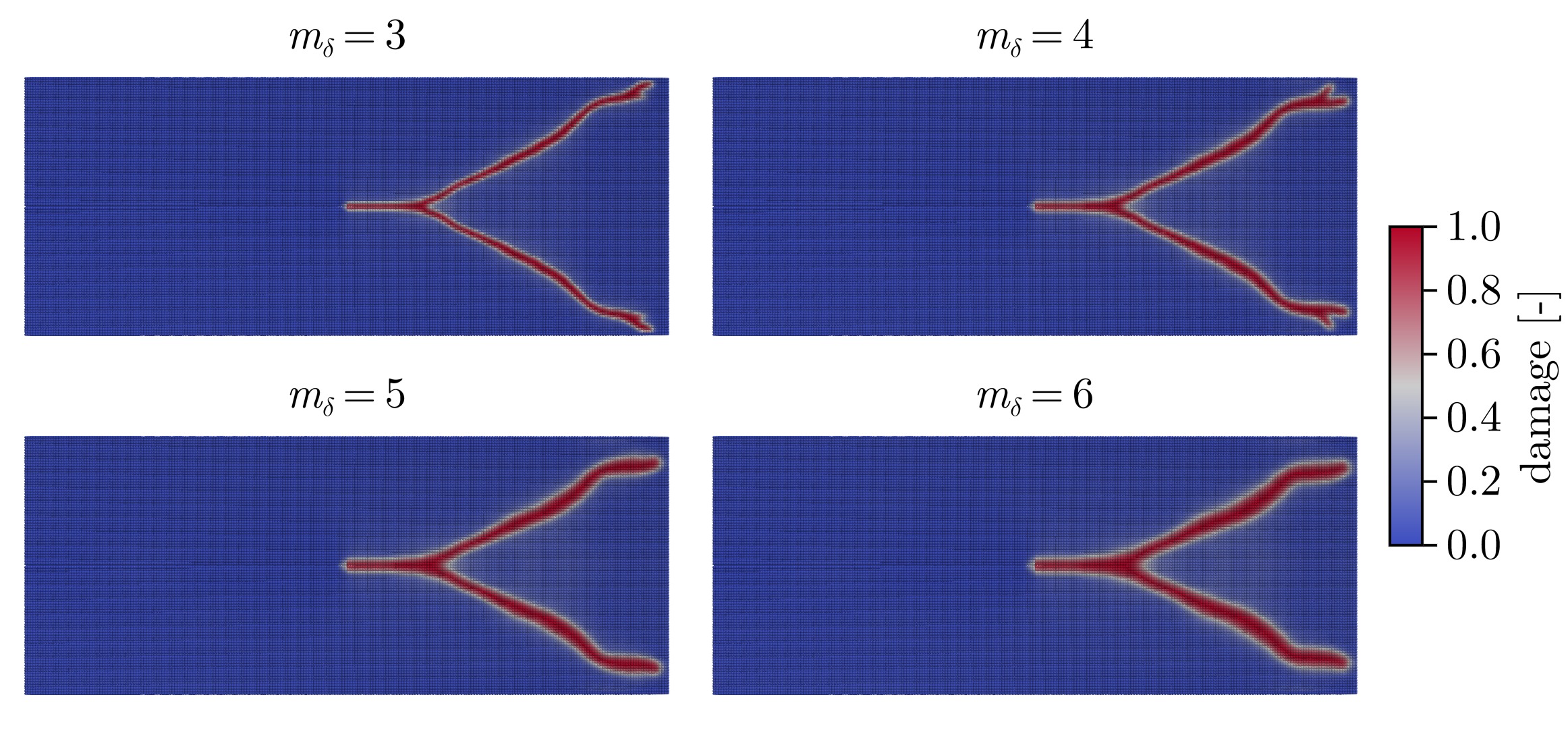}
\caption{Crack branching patterns for different horizon ratios in the BTT for the PFPD model with the linear kernel from \equationref{eq:baqpkernelverification}}
\label{fig:bttbaqppf_k1_c0verification}
\end{figure}

\subsubsection{Results with the linear kernel}

To ensure that the normalization constant is also correct for other kernel functions, the simulations are repeated with the linear kernel
\begin{equation}\label{eq:baqpkernelverification}
\kernel(\Bond) = \frac{3}{\pi \, \horizon^3} \left(1 - \frac{\abs{\Bond}}{\horizon}\right) \eqcomma
\end{equation}
which fulfills the normalization condition from \equationref{eq:kernelintegralnormalization} with $\kernelintegral(\X) = 1$.
The results for this kernel are shown in \autoref{fig:bttbaqppf_k1_c0verification}.
Again, all simulations produce very similar crack paths, and the results are also comparable to those obtained with the cubic B-spline kernel in \autoref{fig:bttbaqppf_k3_c0verification}.
The branching angles and branching points are consistent across both kernels and all horizon ratios, confirming that the normalization constant is correct for different kernel functions.
The obtained crack patterns are in good agreement with the experimental observations reported in the literature~\cite{Bourdin2008,Ha2010}.

\subsubsection{Crack tip velocity}

\begin{figure}[!ht]
\centering
\includegraphics[width=\figsize{0.95\textwidthdissertation}]{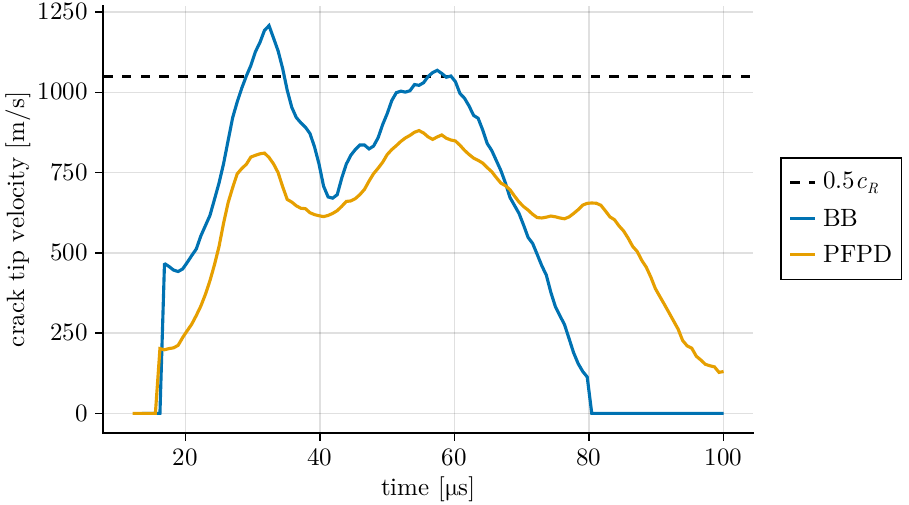}
\caption{Crack tip velocity over the propagation time in the BTT for the PFPD model compared to the bond-based (BB) model, normalized by the half Rayleigh wave speed~$\frac{1}{2}c_R$}
\label{fig:btt_ctv}
\end{figure}

The crack tip velocity over the crack propagation time frame is shown in \autoref{fig:btt_ctv} for the PFPD model with $N_y = 100$, together with results from the bond-based (BB) model for reference.
The crack tip velocity of the PFPD model does not significantly exceed the half Rayleigh wave speed~$\frac{1}{2}c_R$, which is the theoretically expected limiting velocity for crack branching in brittle materials.
The Rayleigh wave speed is approximated as
\begin{equation}
c_R = c_s \, \frac{0.862 + 1.14 \, \poissonsratio}{1 + \poissonsratio} \eqcomma
\end{equation}
with the shear wave speed $c_s = \sqrt{G / \densityref}$ and the shear modulus $G = \youngsmodulus / (2(1 + \poissonsratio))$.
The velocity curve of the PFPD model is smoother compared to the BB results, without the strong peaks observed for the standard formulations.
This observation is attributed to the improved kinematic approximation provided by the correspondence-based framework.

\subsubsection{Discretization convergence}

To verify that the results are consistent across different spatial discretizations, the BTT is simulated with different grid resolutions $N_y \in \{60, 80, 100, 120\}$, corresponding to point spacings $\pointspacing = \frac{2}{5} L_x / N_y$ of approximately $\qty{0.67}{mm}$, $\qty{0.50}{mm}$, $\qty{0.40}{mm}$, and $\qty{0.33}{mm}$, respectively.
All other parameters remain identical to the setup described above, with the cubic B-spline kernel from \equationref{eq:baqpcubicsplinekernel} and a horizon ratio of $\horizonratio = 3$.

\begin{figure}[!ht]
\centering
\includegraphics[width=0.99\textwidth]{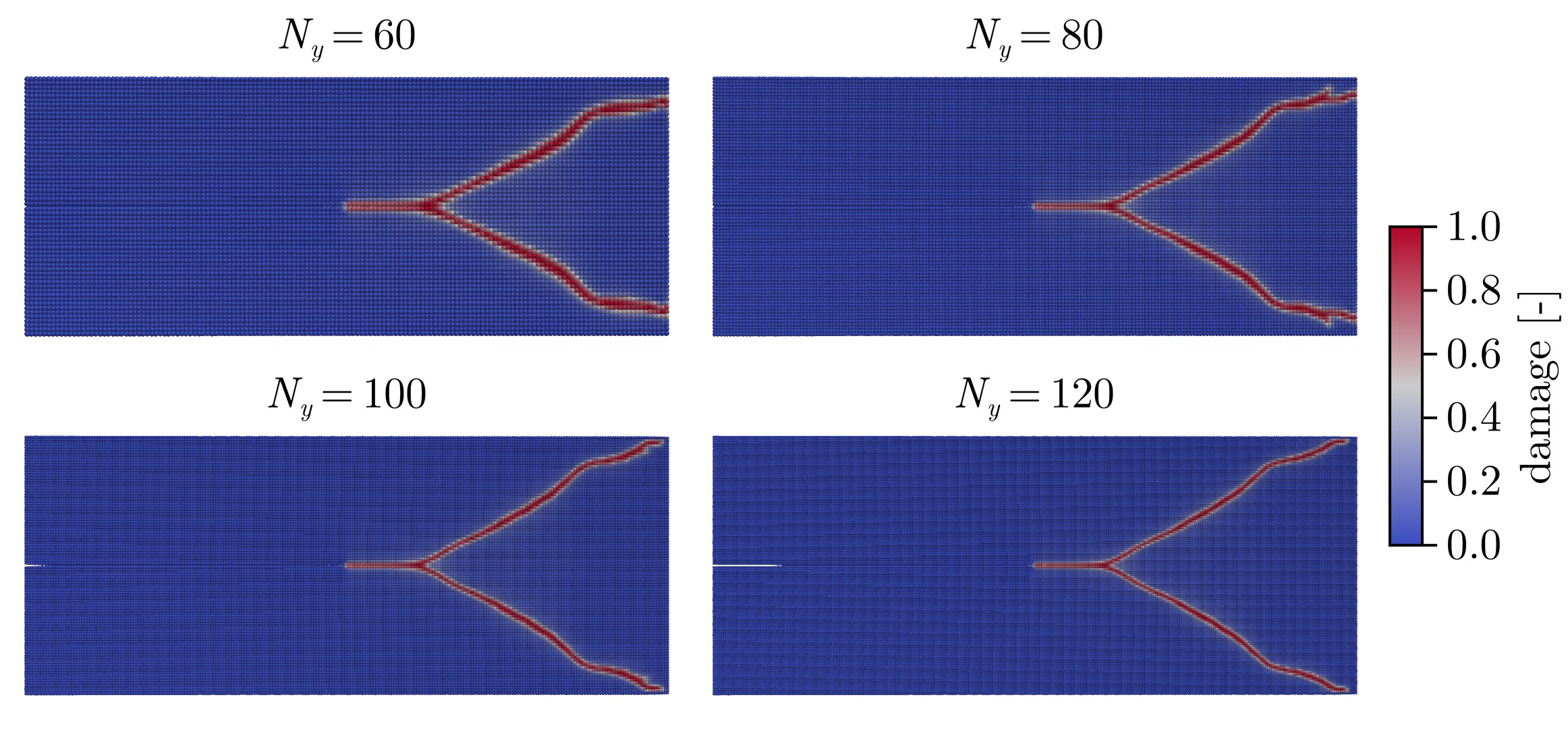}
\caption{Crack branching patterns in the BTT for different discretizations $N_y \in \{60, 80, 100, 120\}$ with the PFPD model}
\label{fig:btt_discretization_convergence}
\end{figure}

In \autoref{fig:btt_discretization_convergence}, the damage fields for the different grid resolutions are shown.
The crack branching pattern is captured for all resolutions, with similar branching angles and branching points across the different discretizations.
As expected, finer discretizations produce narrower damage bands while preserving the same crack topology.
The crack paths are consistent with the results obtained for different horizon ratios in \autoref{fig:bttbaqppf_k3_c0verification} and with phase-field fracture simulations in the literature~\cite{Bourdin2008}, confirming that the PFPD model correctly reproduces the expected fracture behavior under mesh refinement.

\subsection{Kalthoff-Winkler experiment}\label{sec:kalthoffwinkler}

\begin{figure}[ht]
\centering
\includegraphics[width=\figsize{0.7\textwidthdissertation}]{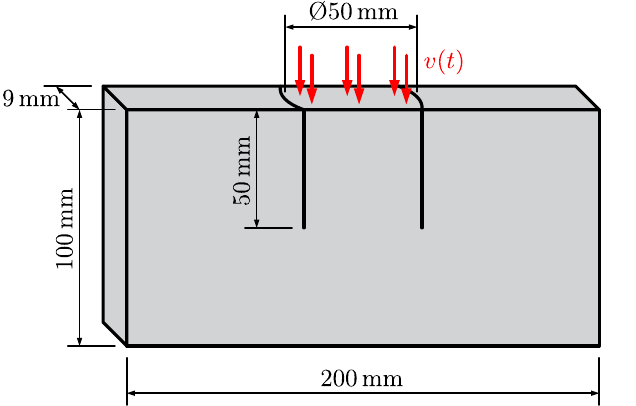}
\caption{Setup of the Kalthoff-Winkler experiment}
\label{fig:kw_setup}
\end{figure}

As a final benchmark, the Kalthoff-Winkler experiment~\cite{Kalthoff2000} is simulated.
This experiment is a well-established test for dynamic fracture under impact loading, where a projectile impacts a pre-notched steel plate, leading to mode~II dominated crack initiation and propagation.
The plate has a width of $\qty{100}{mm}$, a height of $\qty{200}{mm}$, and a thickness of $\qty{9}{mm}$, featuring two parallel pre-notches with a length of $\qty{50}{mm}$ and a cylindrical section diameter of $\qty{50}{mm}$.
The projectile impact is modeled using a Dirichlet boundary condition as in \equationref{eq:mode_i_ii_dirichlet}, with $\Velocity_\mathrm{D}(t,\X) = - v(t) \, \vb{e}_2$ applied to the first five point layers.
The velocity function follows a linear ramp to the impact velocity $v_0$ over a ramp time $t_1$, defined as
\begin{equation}
v(t) = \begin{cases}
v_0 \, \frac{t}{t_1} & \text{if}~t \leq t_1 \eqcomma \\
v_0 & \text{if}~t > t_1 \eqcomma
\end{cases}
\end{equation}
with impact velocity $v_0 = \qty{16.5}{m/s}$ and ramp time $t_1 = \qty{10}{\micro\second}$.
Three layers of boundary points on the opposite side of the impact are defined as no-failure zones.
The plate is discretized with $111 \times 222 \times 10$ material points and a point spacing of $\pointspacing = \qty{0.9}{mm}$.
The material parameters are density $\densityref = \qty{8000}{kg/m^3}$, Young's modulus $\youngsmodulus = \qty{190000}{MPa}$, Poisson's ratio $\poissonsratio = 0.3$, and critical energy release rate $\griffith = \qty{34}{kJ/m^2}$.
The experimental setup is illustrated in \autoref{fig:kw_setup}.

\begin{figure}[!ht]
\centering
\includegraphics[width=0.99\textwidth]{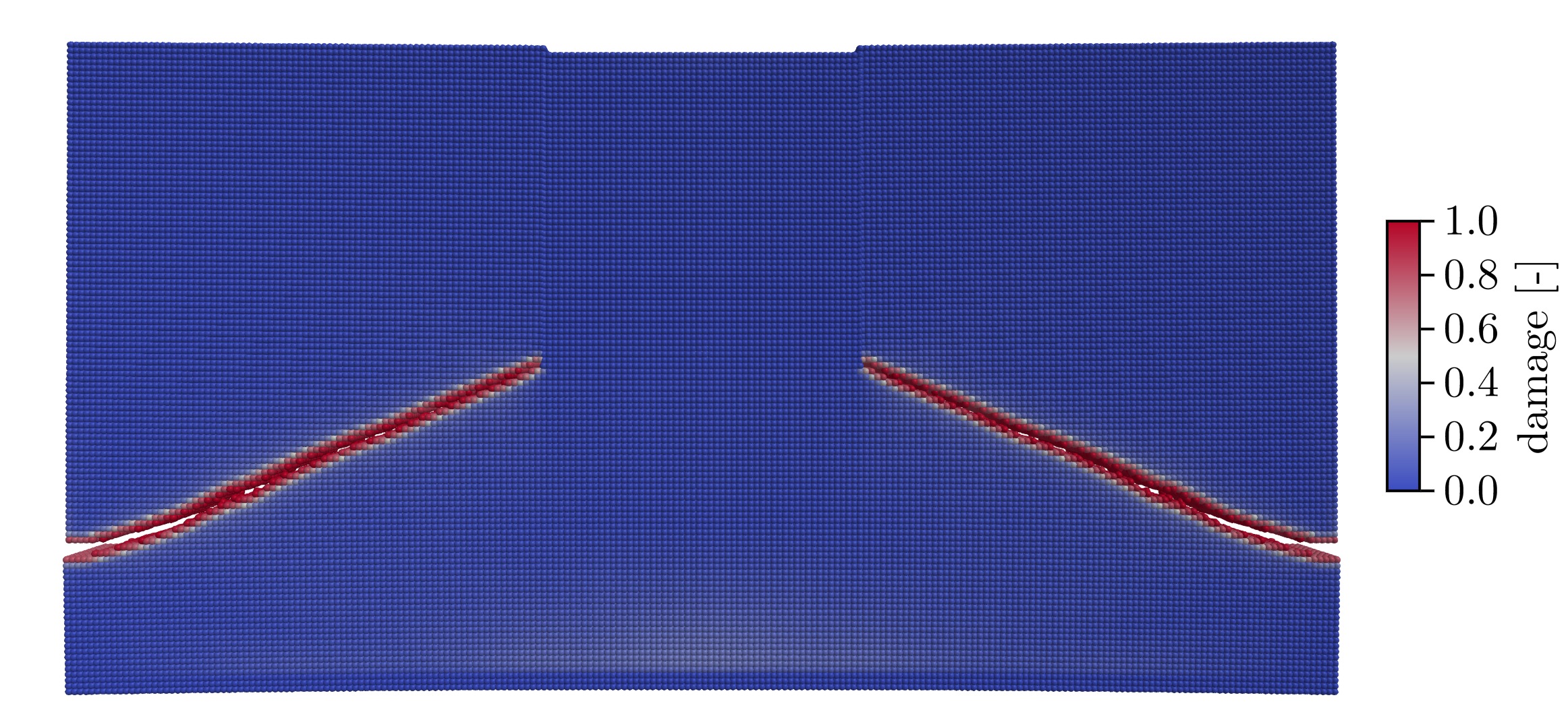}%
\caption{Results of the Kalthoff-Winkler experiment at $t = \qty{0.2}{ms}$ for the PFPD model}
\label{fig:kalthoffwinkler_results_pfpd}
\end{figure}

In \autoref{fig:kalthoffwinkler_results_pfpd}, the damage field of the PFPD model at time $t = \qty{0.2}{ms}$ is shown.
The cracks initiate at both notch tips and propagate at approximately $\qty{70}{\degree}$ to the original notch direction, which is in good agreement with the experimentally observed crack angle~\cite{Kalthoff2000}.

\begin{figure}[!ht]
\centering
\includegraphics[width=\figsize{0.95\textwidthdissertation}]{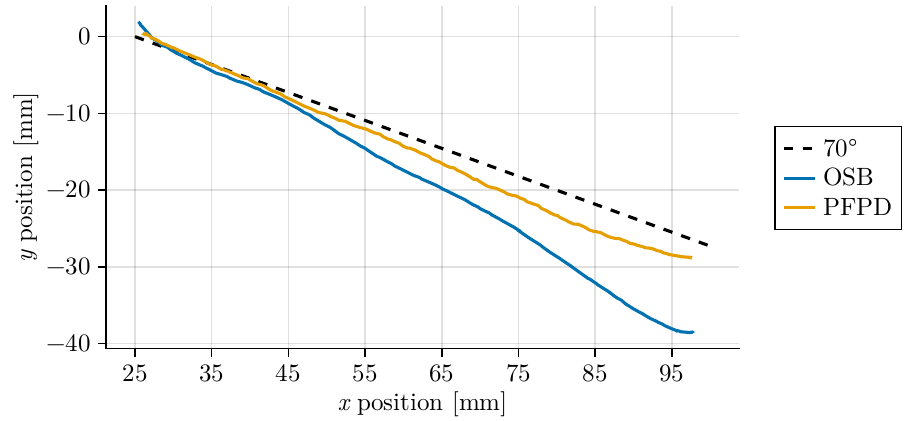}%
\caption{Crack tip trajectory in the Kalthoff-Winkler experiment for the PFPD model compared to the ordinary state-based (OSB) model, with the characteristic $\qty{70}{\degree}$ angle from experiments indicated}
\label{fig:kalthoffwinkler_pfpd_trajectory}
\end{figure}

The crack tip trajectory is shown in \autoref{fig:kalthoffwinkler_pfpd_trajectory}, compared to results obtained with the ordinary-state-based (OSB) peridynamics model.
The PFPD model yields a crack angle closer to the characteristic $\qty{70}{\degree}$ observed in experiments, particularly in the initial phase of crack propagation.
This improved agreement is attributed to the continuous damage evolution in the PFPD framework, which allows for a more gradual and physically consistent crack initiation compared to the abrupt bond deletion in the standard models.

\begin{figure}[!ht]
\centering
\includegraphics[width=\figsize{0.95\textwidthdissertation}]{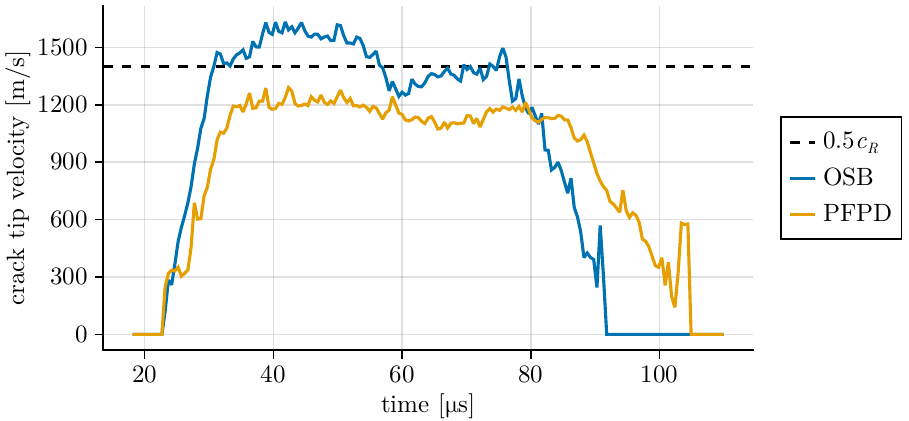}%
\caption{Crack tip velocity in the Kalthoff-Winkler experiment for the PFPD model compared to the ordinary state-based (OSB) model}
\label{fig:kalthoffwinkler_pfpd_vel_vs_time}
\end{figure}

The crack tip velocity over the propagation time frame is shown in \autoref{fig:kalthoffwinkler_pfpd_vel_vs_time}.
The PFPD model yields a smooth velocity profile, with the crack tip velocity remaining below the half Rayleigh wave speed~$\frac{1}{2}c_R$ throughout the simulation.
This is consistent with the expected behavior for the Kalthoff-Winkler experiment, where the fracture is dominated by mode~II loading conditions.
In comparison, the OSB model reaches velocities above~$\frac{1}{2}c_R$ and exhibits a slightly earlier onset of crack propagation (see \autoref{fig:kalthoffwinkler_pfpd_vel_vs_time}).
The crack initiation time of the PFPD model is comparable to that of the BAQP model, further confirming the consistency of the correspondence-based kinematic approximation.

\section{Conclusion}\label{sec:conclusion}

A phase-field peridynamics (PFPD) approach has been developed that embeds a phase-field fracture model into the bond-associated correspondence framework.
Instead of abruptly removing bonds when a critical deformation is exceeded, the approach continuously degrades their energetic contribution through a bond phase-field parameter, thereby avoiding the numerical instabilities inherent to bond deletion in the correspondence formulation.
Additionally, a kinematic degradation function has been introduced to gradually reduce the kinematic contribution of highly damaged bonds, preserving the stability of the shape tensor and the accuracy of the deformation gradient approximation even under large-scale fracture.

An analytical expression for the normalization constant of the damage model has been derived for general spherical kernel functions by requiring that the total crack dissipation equals the Griffith energy release rate per unit crack area.
By reducing the problem to a ratio of two one-dimensional integrals over the unit interval, an exact closed-form expression is obtained that can be evaluated analytically for polynomial kernels.

Numerical examples confirm the capabilities of the PFPD model.
Mode~I and mode~II simulations yield the expected crack paths, with the PFPD model remaining stable in the mode~II case where the standard BAQP model with bond deletion becomes unstable due to shape tensor degradation.
Consistent crack branching patterns in the boundary tension test verify the correctness of the normalization constant across different horizon ratios and kernel functions, with branching angles and branching points in good agreement with experimental observations.
Crack tip velocities do not significantly exceed the half Rayleigh wave speed, consistent with the theoretically expected limiting velocity for crack branching.
In the Kalthoff-Winkler experiment, crack angles of approximately $\qty{70}{\degree}$ are obtained, matching the experimentally observed values, and smoother velocity profiles are achieved compared to standard formulations.

Overall, the proposed framework retains the full accuracy of the bond-associated correspondence formulation while providing a thermodynamically consistent description of fracture that avoids the numerical instabilities associated with bond deletion approaches.
Since bonds are never removed but only degraded, the kinematic approximation remains well-conditioned throughout the simulation, making the PFPD model a robust and versatile tool for dynamic fracture problems.

\section*{Acknowledgments}
The authors gratefully acknowledge the support of the Deutsche Forschungsgemeinschaft (DFG) in the projects \mbox{WE~2525/15-1} and \mbox{WE~2525/15-2} within the Special Priority Program 2256 "Variational Methods for Predicting Complex Phenomena in Engineering Structures and Materials" \#422730790.

\printbibliography

\end{document}